\documentclass[12pt,preprint]{aastex}

\shorttitle{Starbursts and AGNs Mass Accretion}
\shortauthors{WATABE ET AL.}

\begin{document}

\title{Nuclear/Circumnuclear Starbursts and Active Galactic Nuclei Mass Accretion in Seyfert Galaxies}
\author{Yasuyuki Watabe$^{1,2,\ast}$, Nozomu Kawakatu$^3$ and Masatoshi Imanishi$^3$}
\affil{
$^1$Center for Computational Sciences, University of
Tsukuba, Ten-nodai, 1-1-1 Tsukuba, Ibaraki 305-8577, Japan; watabe@ccs.tsukuba.ac.jp\\
$^2$INAF-Osservatorio Astrofisico di Arcetri, Largo Enrico Fermi 5, 50125 Firenze, Italy\\
$^3$National Astronomical Observatory of Japan, 2-21-1 Osawa, Mitaka, Tokyo 181-8588, Japan}
\altaffiltext{$\ast$}{Research Fellow of the Japan Society for the Promotion of Science (JSPS).}

\begin{abstract}
We investigated the correlation between nuclear/circumnuclear starbursts around the active galactic nuclei (AGNs) and the AGN activities for 43 Seyfert galaxies in the CfA and 12 $\mu \rm m$ samples. We found that circumnuclear starburst luminosity as well as nuclear starburst luminosity are positively correlated with AGN luminosity. Moreover, nuclear starburst luminosity is more strongly correlated with the AGN luminosity normalized with AGN Eddington luminosity than is circumnuclear starburst luminosity. This implies that starbursts nearer the AGN could have a greater effect on AGN mass accretion.
We also discuss these results from the viewpoint of the radiation effects from starbursts and sequential starbursts.

\end{abstract}

\keywords{galaxies: active --- galaxies: nuclei --- galaxies: Seyfert --- galaxies: starburst --- infrared: galaxies}

%%%%%%%%%%%%%
%%% introduction %%%
%%%%%%%%%%%%%

\section{INTRODUCTION}

Since the discovery of active galactic nuclei (AGNs) the physical mechanism of AGN fueling remains unresolved. Various fueling mechanisms have been considered thus far, for example, tidal torque driven by major/minor galaxy merger \citep{He89,BH91,MH96,Ta99,SW04}, tidal torque from nonaxisymmetric gravitational potential due to a stellar bar \citep{No88, Sh90, BH92, Kn95, Be96, Fu00}, shock or turbulence in the interstellar medium \citep{Fu98,Mo99,WN99,WN01,Ma02,Wa02,On04}, and radiation drag from a starburst \citep{Um97,Fu97,Um98,Oh99}.

Numerous observations have gradually clarified circumnuclear starburst events (several hundreds of parsecs to a few kiloparsecs from the center) around AGNs. These starbursts often exhibit patchy and ringlike structures \citep{Po89,Wi91,Fo94,Ma94,Mau94,Bu95,Ba95,Le96,Ma96,St96,El98,Kn02,Kn05}. In addition, hidden nuclear starbursts (within a few hundred parsecs from the center) are found both in Seyfert 1 and 2 galaxies \citep{Im02,Im03,RV03} and the nuclear starburst luminosity is positively correlated with the AGN power \citep{IW04}. 

These starburst events could strongly influence the structure and the dynamics of gas through the energy input of multiple supernova explosions \citep{SF76,TI86,NI89,WN02} and the effects of strong radiation \citep{Um97,Um98,OU99,OU01,KU02,WU05,Th05}. 
Especially, for the AGN mass accretion, \citet{WN02} shows the possibility of AGN fueling from the effect of the chaotic disturbance by the nuclear starburst using three-dimensional hydrodynamic simulation. Moreover, \citet{Um97,Um98} have shown that the radiation effects of circumnuclear starburst could cause AGN fueling. 

These theoretical and observational results suggest that some connections exist between the nuclear/circumnuclear starburst and AGN activity. Although a relationship between the nuclear starburst and the AGN has been reported \citep{IW04,Da07}, this remains uncertain regarding the circumnuclear starburst. Moreover, it is not clear whether a specific relationship exists between AGN mass accretion and the starburst at various scales. In this study, we investigated the relationship between the nuclear/circumnuclear starburst and AGN activity by considering both nuclear and the circumnuclear starbursts. 

To investigate nuclear/circumnuclear starbursts, we used the polycyclic aromatic hydrocarbon (PAH) emission feature at 3.3 $\mu$m, 6.2 $\mu$m, 7.7 $\mu$m, and 11.3 $\mu \rm m$. The PAHs are excited by far-UV photons and strong PAH emission are seen in star-forming regions, whereas a pure AGN shows only a featureless spectrum with virtually no PAH emission. Also, since the dust extinction is much lower for these $3.3 - 11.3 \mu \rm{m}$ PAH emission \citep[$\lesssim 0.06 A_{\rm V}$;][]{RL85,Lu96}, we can quantitatively estimate modestly obscured ($A_{\rm V} < 15$  mag) star-formation activity by using these emission lines. Throughout the paper, we adopted $H_{0} = 80\, \rm km s^{-1} Mpc^{-1}$, $\Omega_{M}=0.3$, and $\Omega_{\Lambda}=0.7$.

%%%%%%%%%%%%%
%%% sample data %%%
%%%%%%%%%%%%%

\section{SAMPLE DATA}
 
We studied 21 Seyfert 1 galaxies and 22 Seyfert 2 galaxies in the CfA \citep{HB92} and 12 $\mu \rm m$ \citep{Ru93} samples selected based on their host-galaxy magnitudes and {\it IRAS} 12 $\mu$m fluxes, respectively. These samples are not expected to be biased toward or against the presence of nuclear starbursts. Our sample was selected from \citet{IW04} data set with both the 3.3 $\mu$m PAH luminosity and the nuclear N-band luminosity. In their sample, Seyfert galaxies at $z = 0.008-0.035$ were selected to investigate the nuclear starburst by ground-based spectroscopy using a $1"$-$2"$ slit, where $1"$ corresponds to a physical scale of 150 ($z = 0.008$) to 650 pc ($z=0.035$). Also, to be best observable from Mauna Kea, Hawaii, the declinations of Seyfert galaxies are limited to being larger than $-30^\circ$. Owing to the telescope limit of the IRTF 3 m telescope, a restriction of declination of less than $68^\circ$ is also applied. For their sample, there are no obvious biases.

Among 43 Seyfert galaxies, we estimated the supermassive black hole (SMBH) mass to investigate the mass accretion rate normalized with the AGN Eddington mass accretion rate for 25 objects. To these objects, we estimated the 6.2 $\mu$m, 7.7 $\mu$m, and 11.3 $\mu$m PAH emission for 13 objects further. Also, we collected the X-ray luminosity from the data in the literature for 15 objects. Since our sample included some upper limit data, we used statistical techniques applicable to a sample where upper limit data are present. 

\begin{figure}
\epsscale{.80}
\plotone{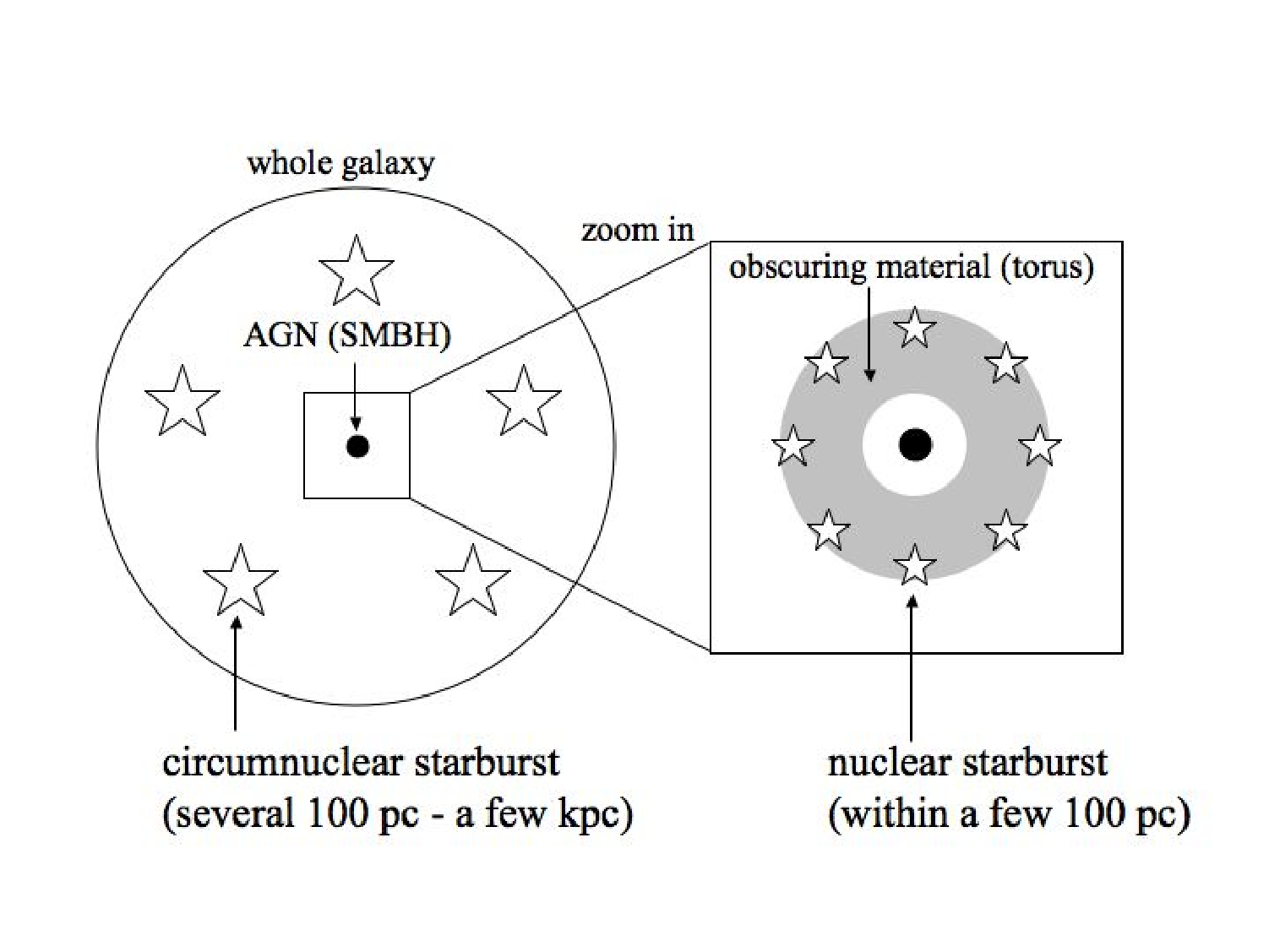}
\caption{The schematic view of the nuclear/circumnuclear starbursts. The nuclear starburst (within a few 100 pc) and the circumnuclear starburst (several 100 pc - a few kpc) are distributed around AGN.}
\label{fig1}
\end{figure}

We focused on both nuclear starbursts, which exist within a few hundred parsecs from the center, and circumnuclear starbursts in the entire host galaxy region. Figure \ref{fig1} shows the schematic view of the nuclear/circumnuclear starbursts. 
We can consider the 3.3 $\mu$m PAH emission, which were derived through ground-based spectroscopy using the narrow slit, as a nuclear starburst. Also, since a circumnuclear starburst is typically much higher than a nuclear starburst (see section 4.2), the 6.2 $\mu$m, 7.7 $\mu$m, and 11.3 $\mu$m PAH emission, which were observed in the whole host galaxy region, represent a circumnuclear starburst. In detail, we described in section 3.1 and 3.2.

%%%%%%%%%%%%%%%%%%%%%%%%
%%% Estimation of physical properties %%%
%%%%%%%%%%%%%%%%%%%%%%%%

\section{ESTIMATION OF PHYSICAL PROPERTIES}

\subsection{Nuclear Starburst}

PAH molecules near the AGN can be destroyed by strong X-ray radiation from the AGN \citep{Vo92,Si04}. However, if the PAH molecules are sufficiently shielded by a substantial column density of X-ray-absorbing gas, PAHs can survive even if they exist near the AGN. In fact, ample gas is believed to exist that could be related to the obscuring material (the dusty torus in the context of the AGN unified model) around the AGN. In addition, the gravitational stability parameter $\mathcal{Q}$ \citep{To64} decreases with the radius of the obscuring material, whose mass is notably smaller than that of the SMBH \citep{Im03}. The gravitational collapse of molecular gas can therefore occur more easily in the outer part of the obscuring material. Thus, in the outer part of the obscuring material, a starburst could exist near the AGN and PAHs could survive. In fact, from the estimation of the surface brightness values of the 3.3 $\mu$m PAH emission \citep{IA04,IW04}, which were derived through ground-based spectroscopy using the $1"$-$2"$ slit, this emission should come from nuclear starbursts in the gas-rich region, which may be the obscuring material around AGN. 

Therefore, we can use their 3.3 $\mu$m PAH emission data, which were observed by ground-based spectroscopy using the narrow slit,  as the indicators of nuclear starbursts. Although the circumnuclear starburst, which often exhibit ring-like structures with a radius of kiloparsecs order from the center, along the slit direction could be also included, the fraction of the circumnuclear starburst emission inside this thin slit is negligible compared to the whole. Thus, the 3.3 $\mu$m PAH emission could be regarded as a practical probe for nuclear starbursts in both Seyfert 1 and Seyfert 2 galaxies.

In \citet{Im03} and \citet{IW04}, only upper limits for the nuclear
3.3 $\mu$m PAH fluxes are available in more than half of the observed
Seyfert galaxies. Additionally, the observed Seyfert galaxies do not comprise 
a complete sample. To increase the number and fraction of Seyfert galaxies with detectable nuclear 3.3 $\mu$m PAH emission, we newly observed eight sources, 
which are shown in Table \ref{tbl-1}.  
Among them, six sources (NGC 931, F03450+0055, NGC 262, NGC 513,
MCG-3-58-7, and Mrk 993) showed possible signs of the 3.3 $\mu$m PAH
emission in our previously obtained infrared 3--4 $\mu$m spectra 
\citep{Im03,IW04}. Therefore, 
we decided to re-observe these sources with longer exposure to check if
the signs are real or not. 
The two sources (MCG-2-8-39 and 0152+06) are newly observed
this time to increase the number of observed sources. We showed these observations in Appendix.

\subsection{Circumnuclear Starburst}

To estimate the absolute magnitudes of circumnuclear starburst 
activity in Seyfert galaxies, we analyzed archival 
infrared data obtained with the Infrared Spectrograph (IRS) 
\citep{Ho04} onboard the Spitzer Space Telescope 
\citep{We04} through the program PID 3269 
(PI = Gallimore).  
We focused on the infrared spectra of Short-Low 2 (SL2; 5.2--7.7 $\mu$m) 
and 1 (SL1; 7.4--14.5 $\mu$m), to estimate the fluxes of 
PAH-emission features at $\lambda_{\rm rest}$ = 6.2 $\mu$m, 7.7 $\mu$m, 
and 11.3 $\mu$m in the rest frame.

We used the latest pipeline-processed data products at the time of our analysis (S11--14, pbcd files). 
The observations were performed with the slit-scan mode, and the entire 
host galaxy regions of the Seyfert galaxies were covered.
We summed all the signals and extracted the spectra after subtracting 
the background emission in a standard manner.  
Thus, the resulting spectra should reflect the emission from the 
entire regions of the Seyfert galaxies. 
Wavelength and flux calibrations for the SL data were made on the basis of the
files of the {\it Spitzer} pipeline-processed data, 
termed ``b0\_wavsamp.tbl'' and ``b0\_fluxcon.tbl'', respectively.
For SL1 spectra, data at $\lambda_{\rm obs}$ $>$ 14.5 $\mu$m in the 
observed frame are invalid (Infrared Spectrograph Data Handbook 
Version 1.0) and so were removed.
For some faint sources, appropriate spectral binning was applied to 
reduce the scatter of data points and to achieve more reliable 
measurements of the PAH fluxes.

The circumnuclear starburst luminosity is typically much higher than the nuclear starburst luminosity \citep[][and see section 4.2]{Im03,IW04}. Thus, we can consider that the 6.2 $\mu$m, 7.7 $\mu$m, and 11.3 $\mu$m PAH emission, which were observed in the whole host galaxy regions, are good indicators of a circumnuclear starburst even if these PAH luminosity contain a small contribution from nuclear starbursts.

\subsection{AGN Luminosity}

For AGN activity, we used the nuclear N-band ($10.5 \mu$m) luminosity measured with a $1.5"$ aperture \citep{Go04} since the contribution of starbursts for the nuclear N-band luminosity is small \citep{Go04}. However, nuclear N-band luminosity represents the reemission of hot dust illuminated by the AGN. So it may contain some uncertainties due to the dust distribution and the covering factor of the dusty torus around the AGN. To examine whether the nuclear N-band luminosity represents AGN activity in the sample data, we also used absorption-corrected hard (2-10 keV) X-ray luminosity, which is regarded as the AGN intrinsic luminosity for the Compton-thin AGN by using the data in the literature (see references in Table \ref{tbl-3}).

\subsection{Black Hole Mass}

Next, we estimated the SMBH mass, $M_{\rm BH}$, to investigate the mass accretion rate normalized with the AGN Eddington mass accretion rate. To estimate SMBH masses of Seyfert 1 galaxies, we used the following method \citep[in detail, see][]{Ka07}. 
The method for estimating the mass of a SMBH is based on the assumption that the motion of ionized gas clouds moving around the SMBH is dominated by the gravitational force and the clouds within the broad-line region (BLR) are virialized \citep[e.g.,][]{PW99}. Adopting an empirical relationship \citep{Ka00} between the size of the BLR and the rest-frame optical continuum luminosity, $\lambda L_{\lambda} (5000 \AA)$ and reverberation mapping, we can obtain the following formula:

\begin{equation}
M_{\rm BH} = 4.9^{+0.4}_{-0.3} \times 10^{6} 
\left[ \frac{\lambda L_{\lambda} (5100 \rm \AA)}{10^{44}\, \rm{ergs\, s^{-1}}} \right]^{0.70\pm0.033} \left( \frac{v_{\rm FWHM}}{10^{3}\, \rm{km\, s^{-1}}} \right)^{2} M_{\odot}.
\end{equation}
We can use this equation for estimating the mass of SMBHs in Seyfert 1 galaxies. 

We cannot use the above method for Seyfert 2 galaxies because no broad emission components exist. So we used the data of SMBH mass \citep{BG07}, which is estimated by SMBH mass and the stellar velocity dispersion, $\sigma_{\ast}$, relation \citep{Tr02}, 
\begin{equation}
M_{\rm BH} = 10^{8.13} \left( \frac{\sigma_{\ast}}{200 \rm{km s^{-1}}} \right)^{4.02} M_{\odot}.
\end{equation}

%%%%%%%%%%
%%% results %%%
%%%%%%%%%%

\section{RESULTS}

\subsection{Nuclear 3.3 $\mu$m PAH Emission}

\begin{figure}
\epsscale{.40}
\plotone{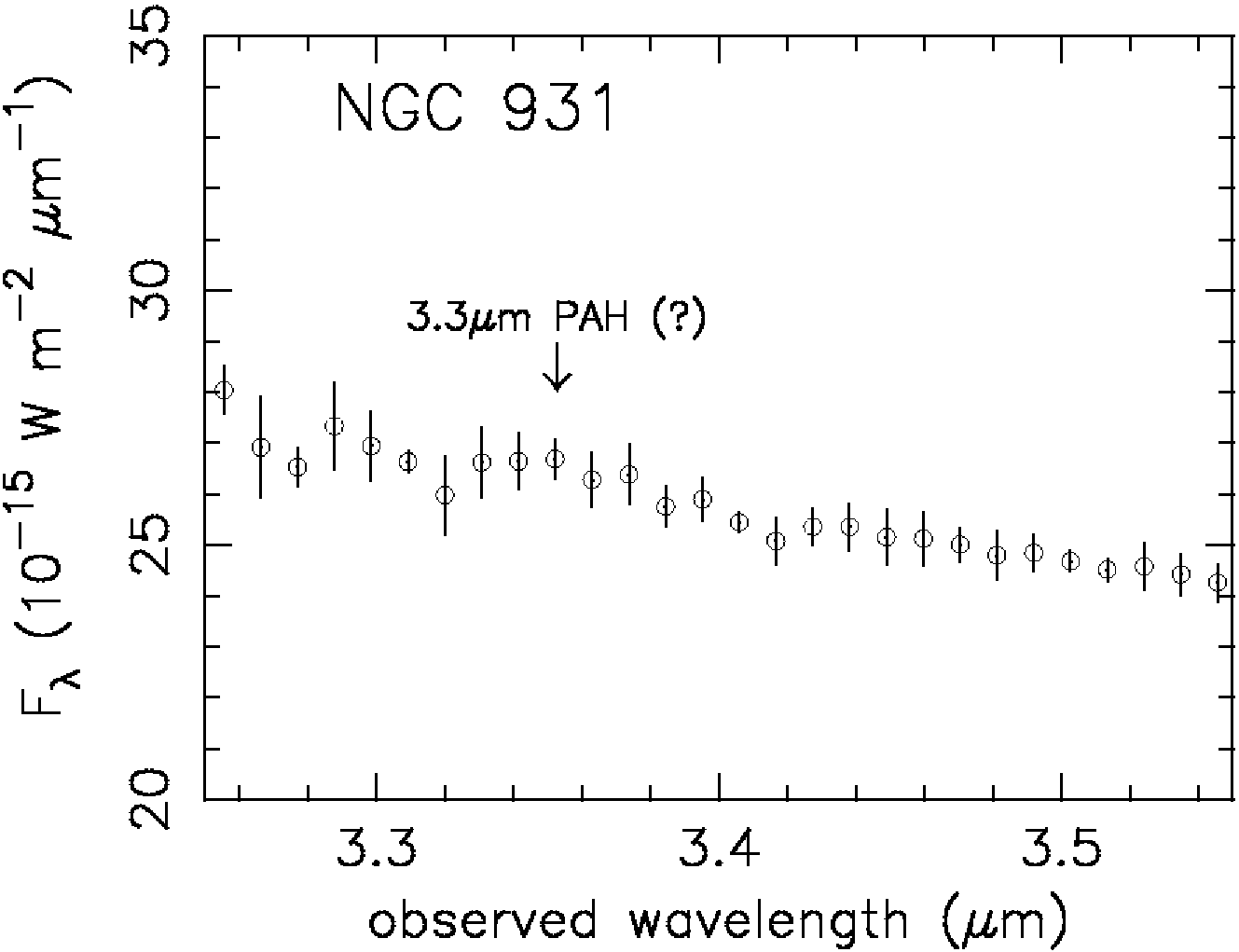} %N931
\plotone{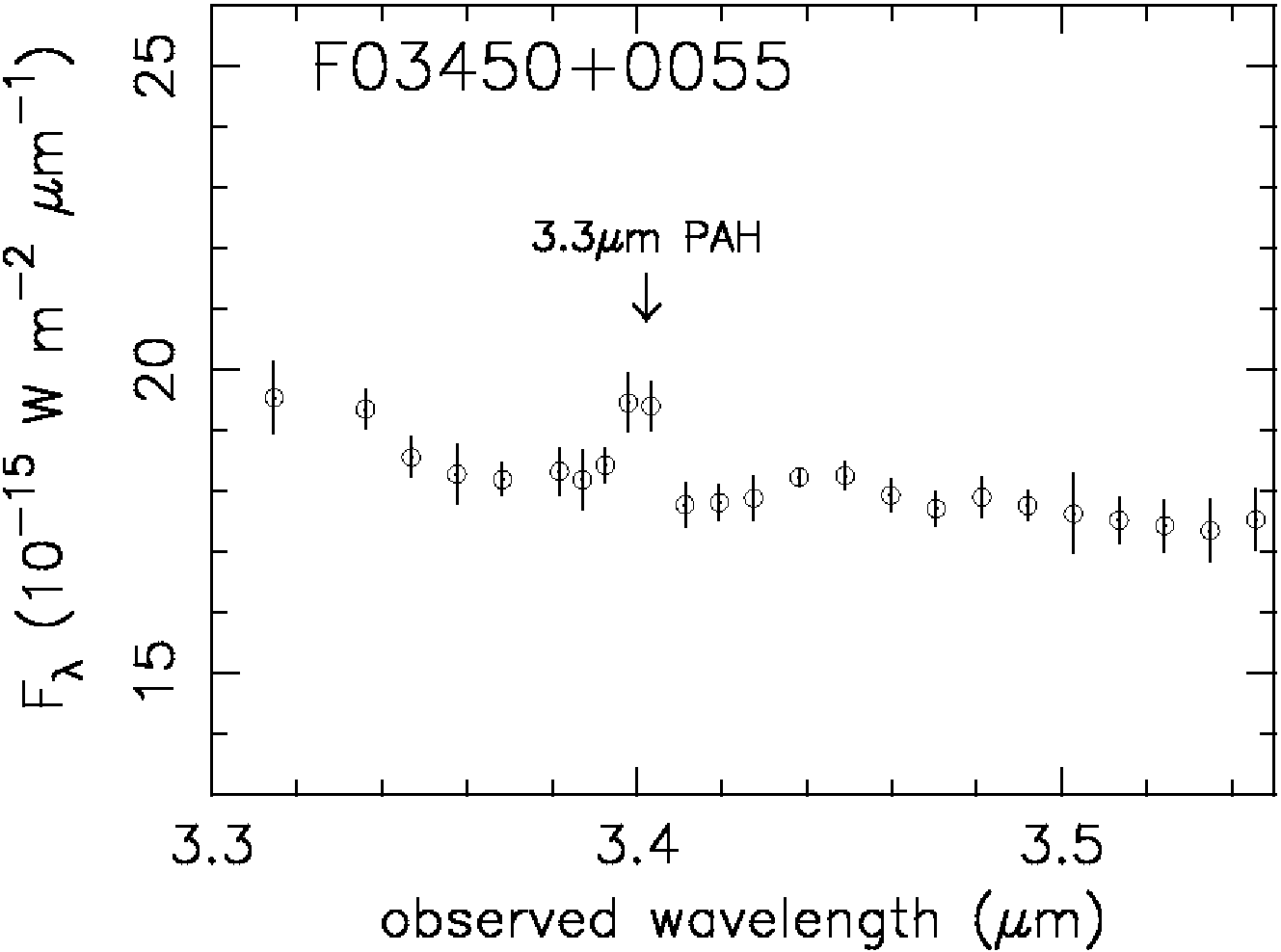} \\%F03450
\plotone{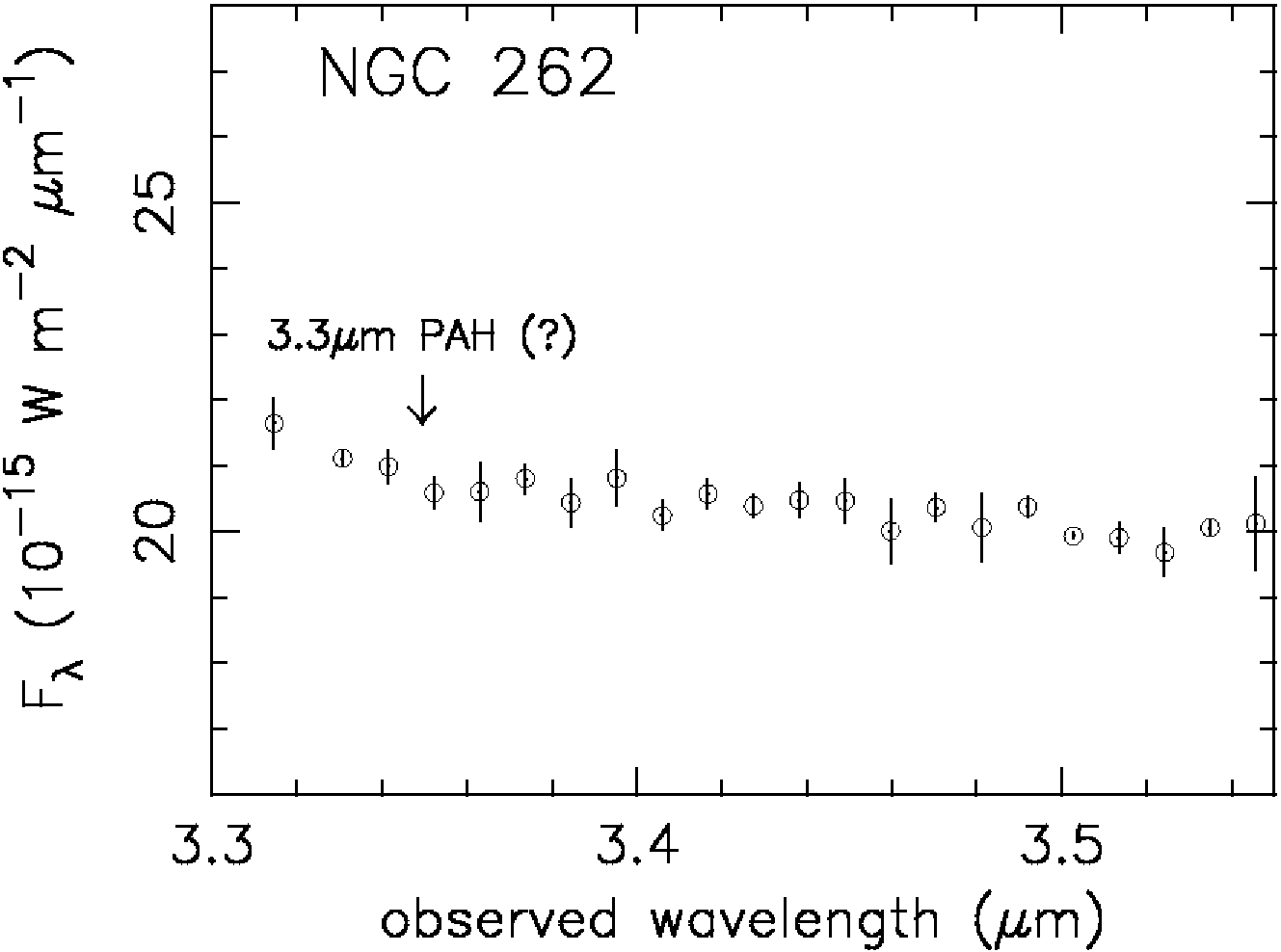} %N262
\plotone{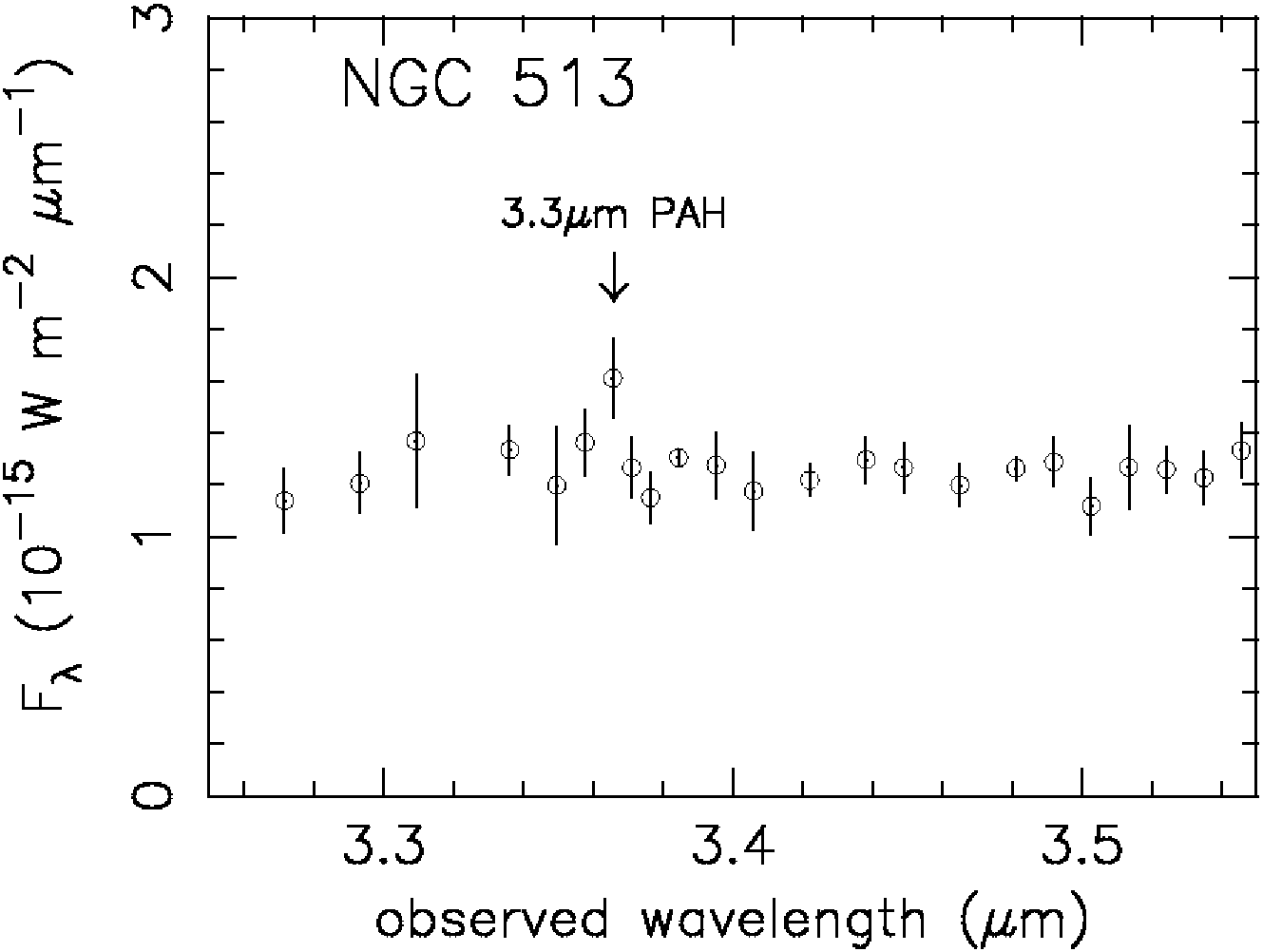} \\%N513
\plotone{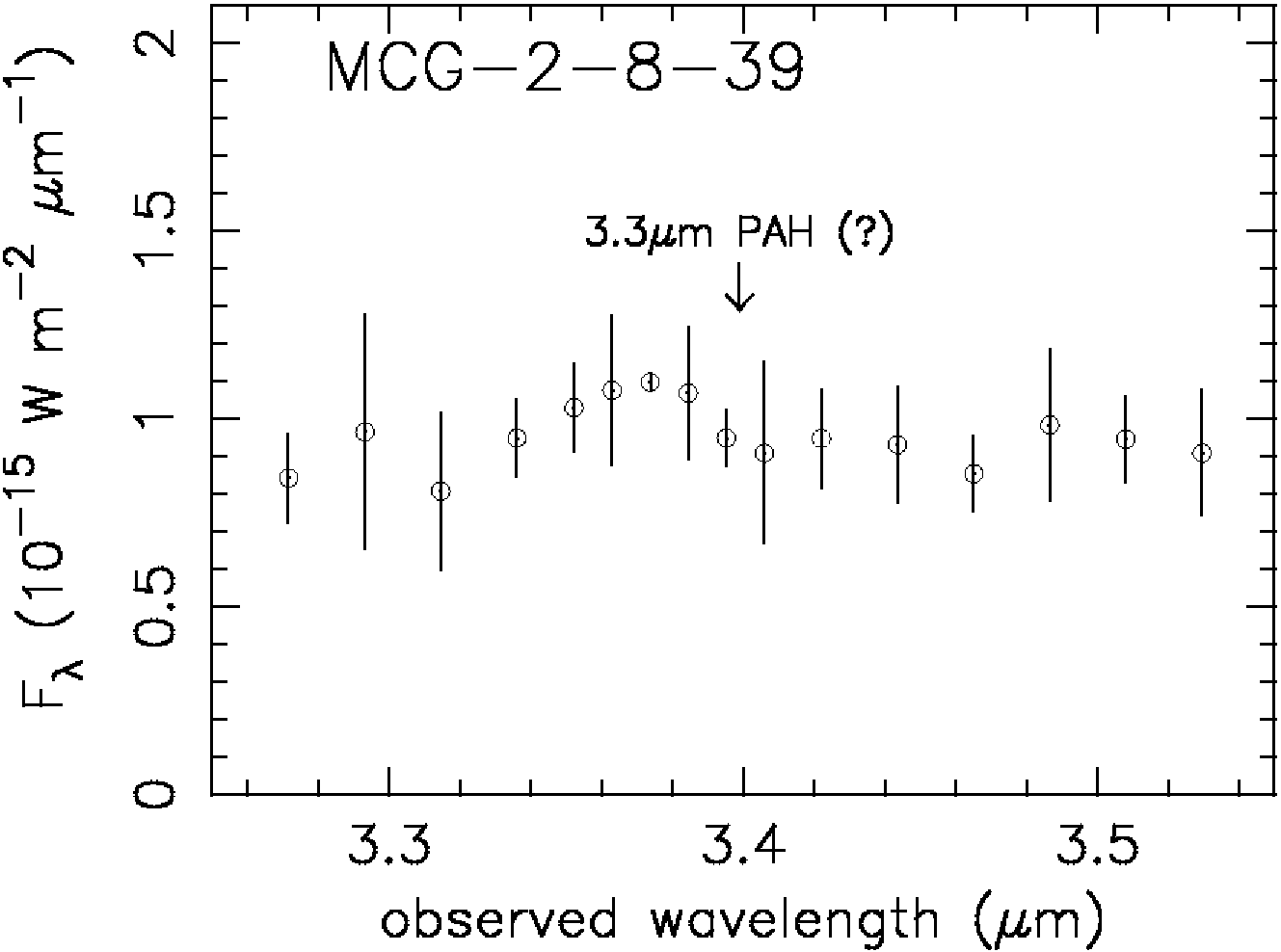} %MCG2839
\plotone{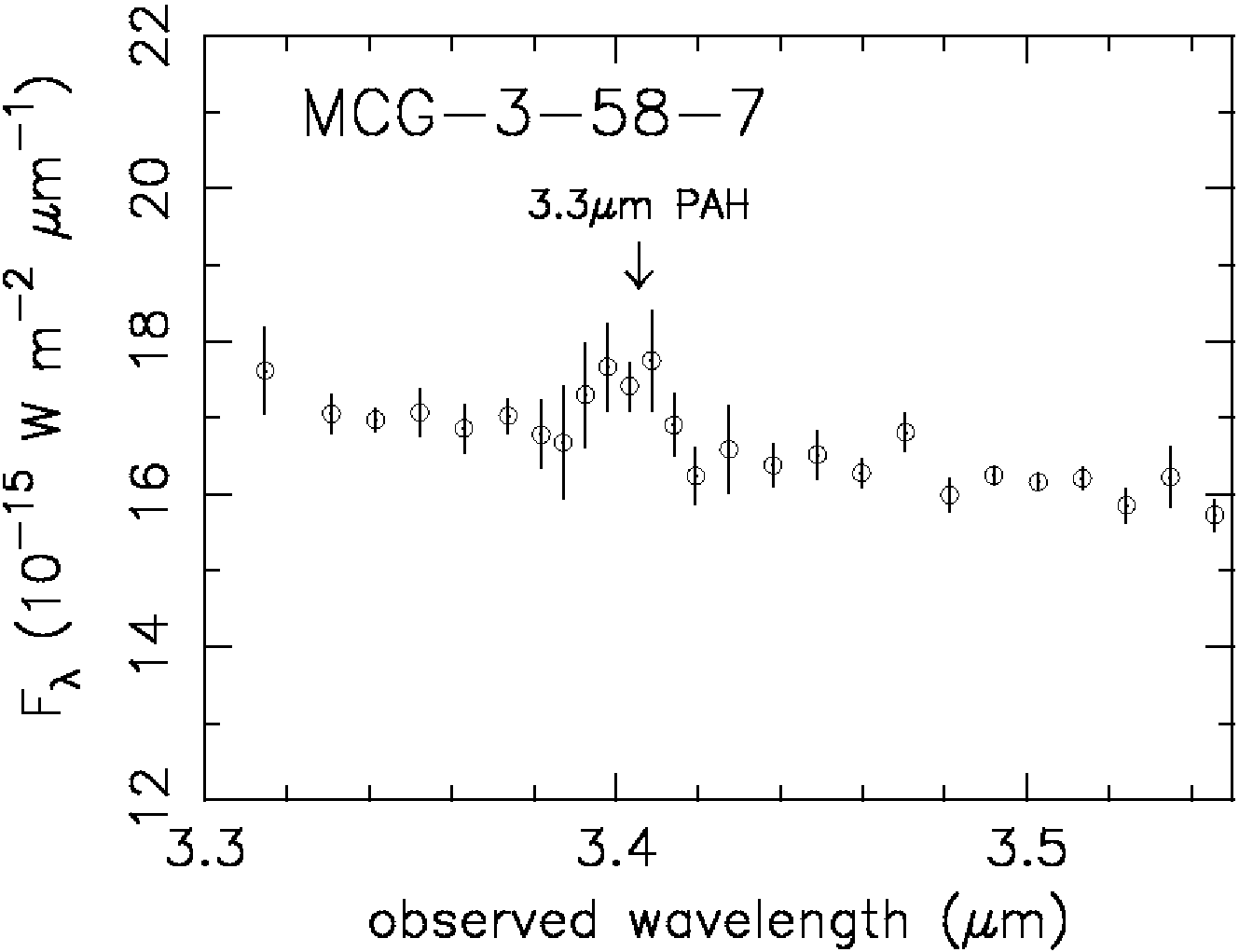} \\%HR3587
\plotone{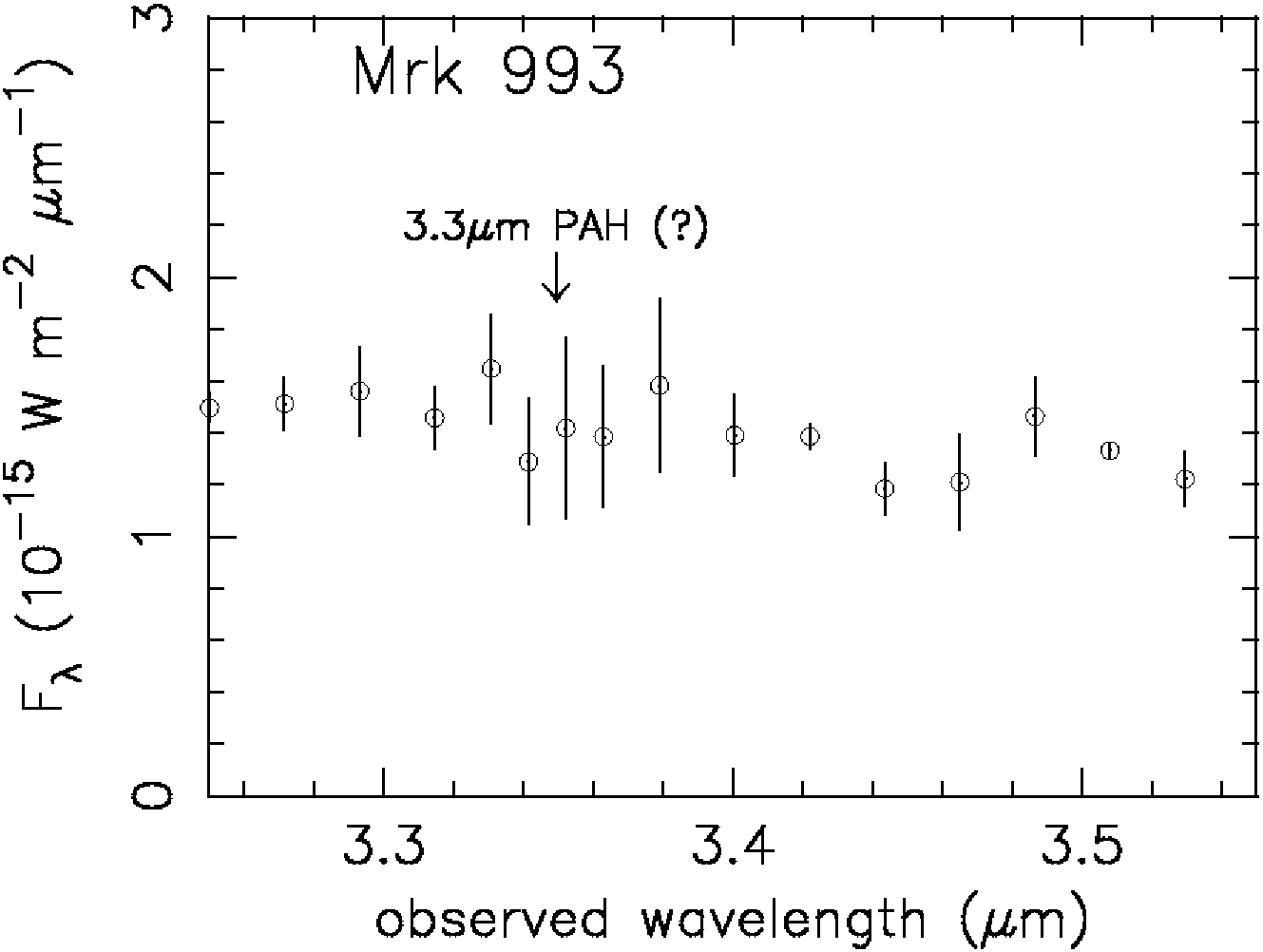} %Mrk993
\plotone{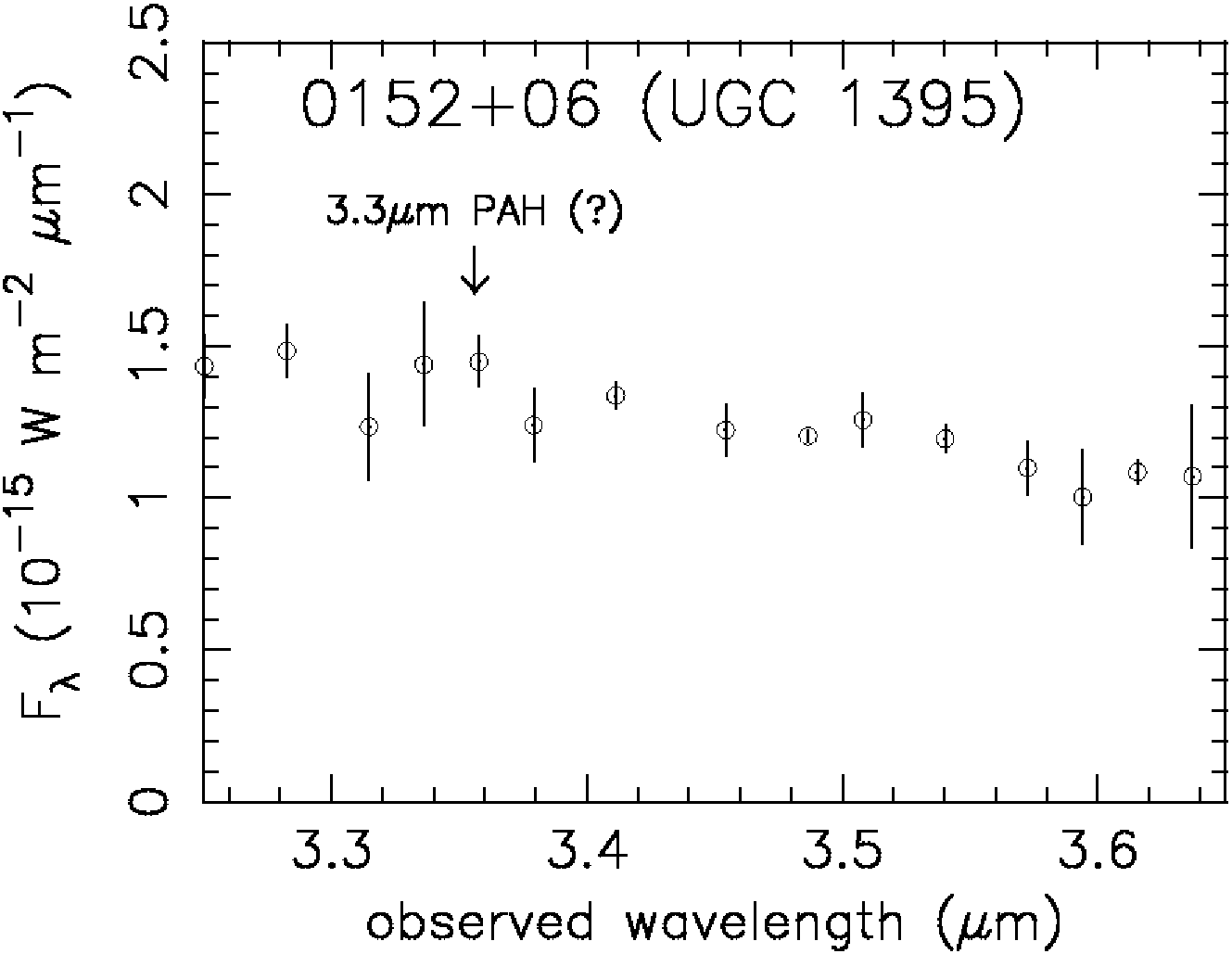} \\%Q152
\caption{
Zoom-in spectra around the redshifted 3.3 $\mu$m PAH emission feature 
(lower arrows) of the observed eight Seyfert galaxies. 
The horizontal and vertical axes are the observed wavelength in $\mu$m and
F$_{\lambda}$ in 10$^{-15}$ W m$^{-2}$ $\mu$m$^{-1}$, respectively.
}
\label{fig2}
\end{figure}

Figure \ref{fig2} represents the zoom-in spectra around the 3.3 $\mu$m PAH emission
feature of the observed Seyfert galaxies.
Thanks to the excellent observing conditions and longer exposure, the 
flux excesses at the expected wavelength of the 3.3 $\mu$m PAH emission
feature at (1 + $z$) $\times$ 3.3 $\mu$m are now recognizable in the
three sources (F03450+0055, NGC 513, and MCG-3-58-7), marked with 
``3.3 $\mu$m PAH'' in Fig. \ref{fig2}.  
For these sources, the fluxes, luminosities, and rest-frame equivalent
widths (EW$_{\rm 3.3PAH}$) of the 3.3 $\mu$m PAH emission were estimated
using Gaussian fits. 
For the remaining five sources with no clear 3.3 $\mu$m PAH detection 
(marked with ``3.3 $\mu$m PAH (?)'' in Fig. \ref{fig2}), we estimated the upper
limits of the PAH strengths, by adopting the lowest plausible continuum
levels and assuming the profile of type-1 sources in \citet{To91} as 
a 3.3 $\mu$m PAH template.   
Table \ref{tbl-2} summarizes the results.

At $\lambda_{\rm rest}$ $\sim$ 3.3 $\mu$m in the rest frame, Pf$\delta$
(3.30 $\mu$m) emission is present, superposed on the 3.3 $\mu$m PAH
emission.  
The relative contribution from this Pf$\delta$ emission line is expected
to be high in Seyfert 1 galaxies, because broad emission line components
are unobscured. 
However, the equivalent width of the Pf$\delta$ line is expected to be 
$\sim$0.3 nm for a typical Seyfert 1 galaxy \citep{ID06}, 
and even lower in a Seyfert 2 galaxy.
This equivalent width value is much lower than the observed values or
upper limits in Table \ref{tbl-2}. 
We thus ascribe the flux excess at $\lambda_{\rm rest}$ = 3.3 $\mu$m
mostly to the 3.3 $\mu$m PAH emission feature. 

Among the six previously observed Seyfert galaxies, the detected 
3.3 $\mu$m PAH fluxes or their upper limits are lower than previously
derived upper limits \citep{Im03,IW04}, except NGC 931. 
For NGC 931, our new upper limit of the 3.3 $\mu$m PAH flux is
$\sim$50\% higher than our previous estimate \citep{IW04}, possibly
because our new estimate is a conservative one, derived based on the
lowest plausible continuum level. 
The 3--4 $\mu$m continuum flux levels are also similar within $\sim$50\%
between the spectra in this paper and in our previous papers
\citep{Im03,IW04}, except NGC 513. 
The flux of NGC 513 in this paper is more than 1 mag fainter than 
in \citet{Im03}, possibly because of narrower slit employed and/or 
the flux variation of the 3--4 $\mu$m continuum emission, originating in 
AGN-heated hot dust.

\subsection{6.2 $\mu$m, 7.7 $\mu$m, and 11.3 $\mu$m PAH Emission}

\begin{figure}
\epsscale{.40}
\plotone{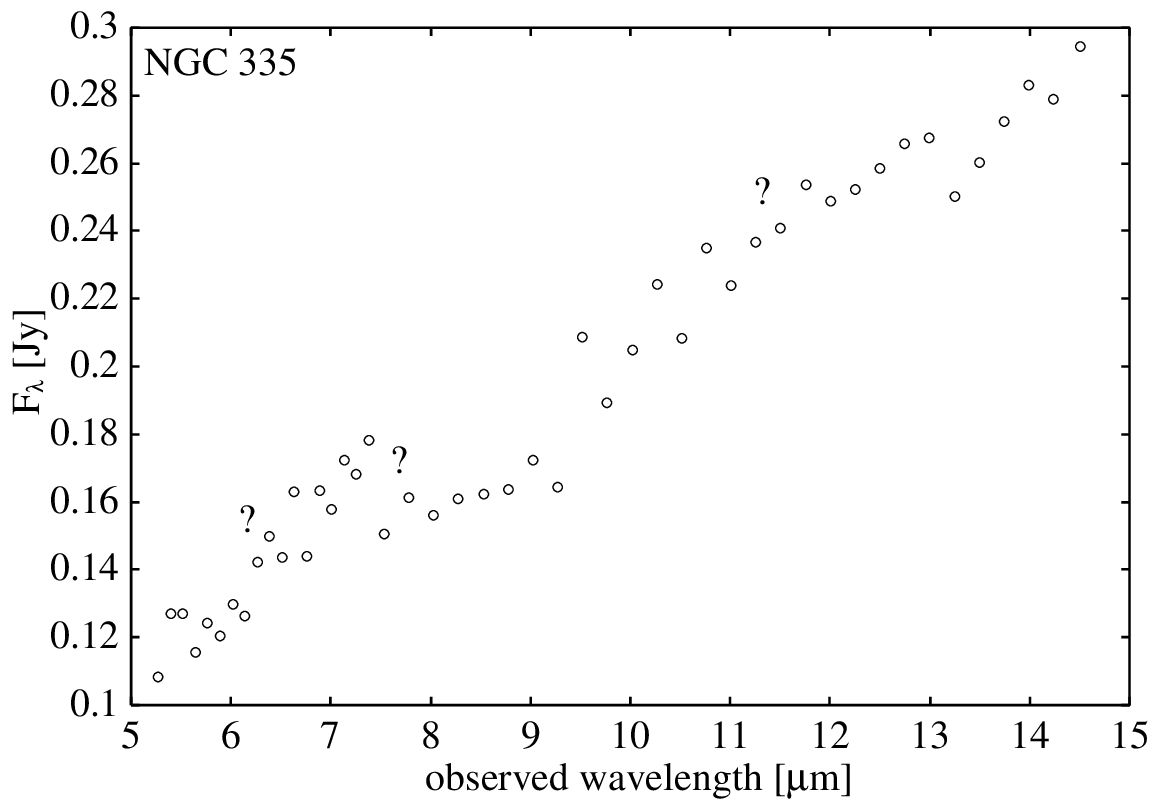}%335b4
\plotone{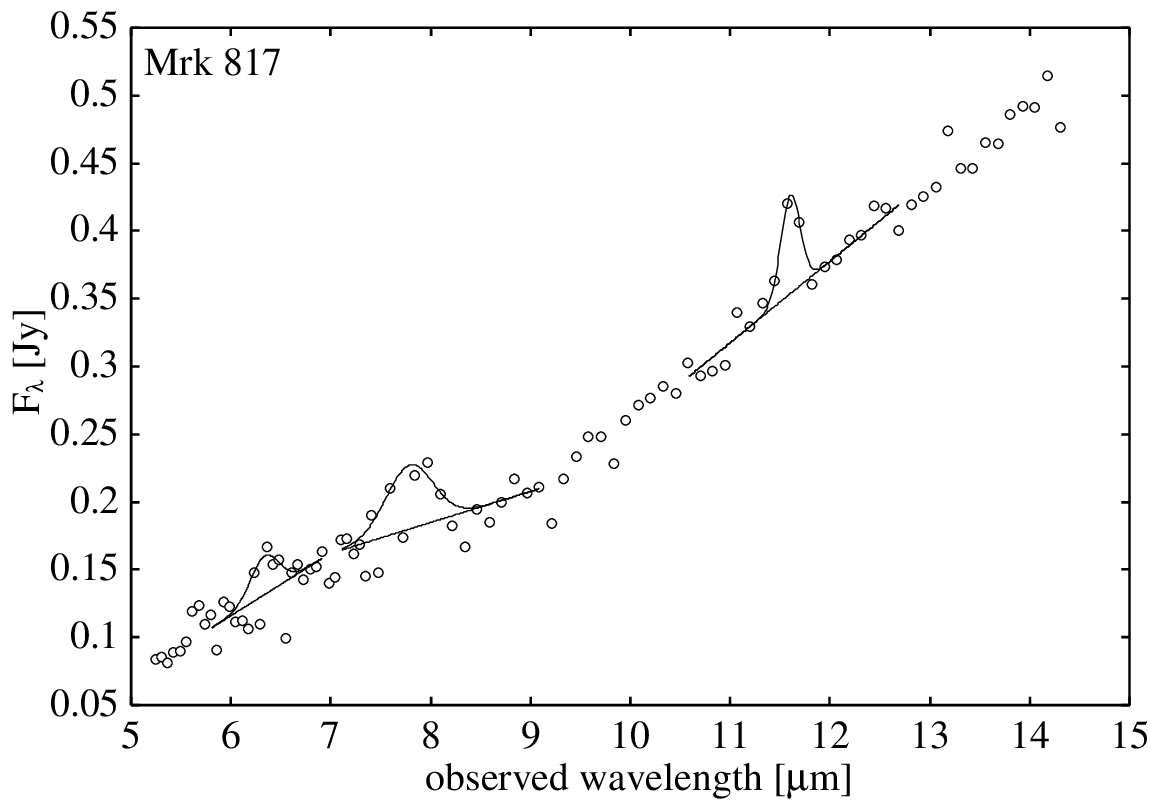}\\%817
\plotone{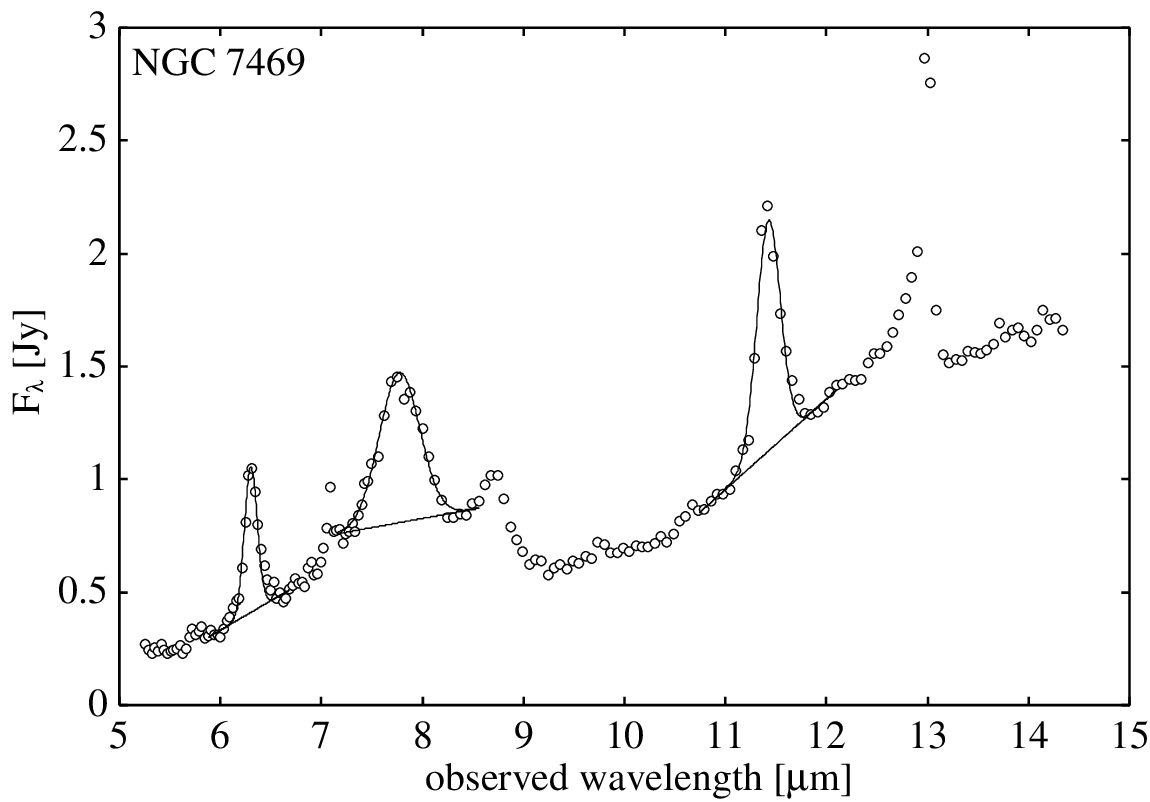}%7469
\plotone{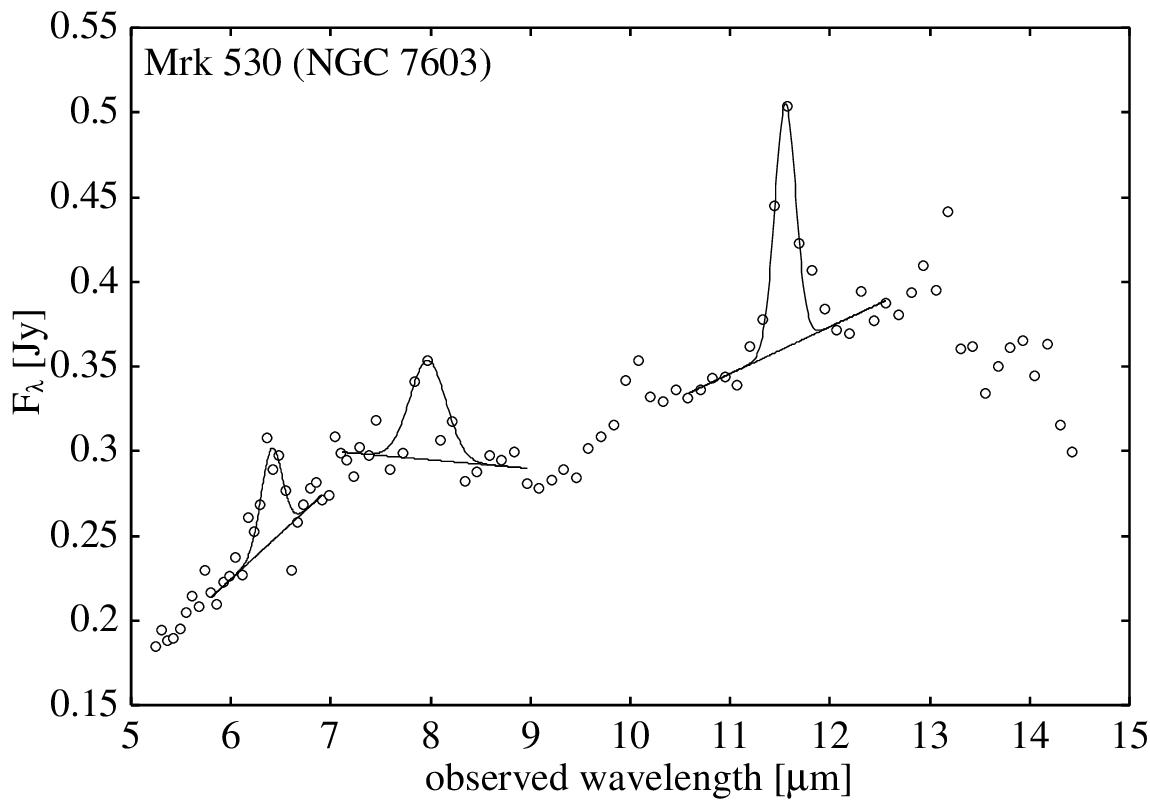}\\%530_7603
\plotone{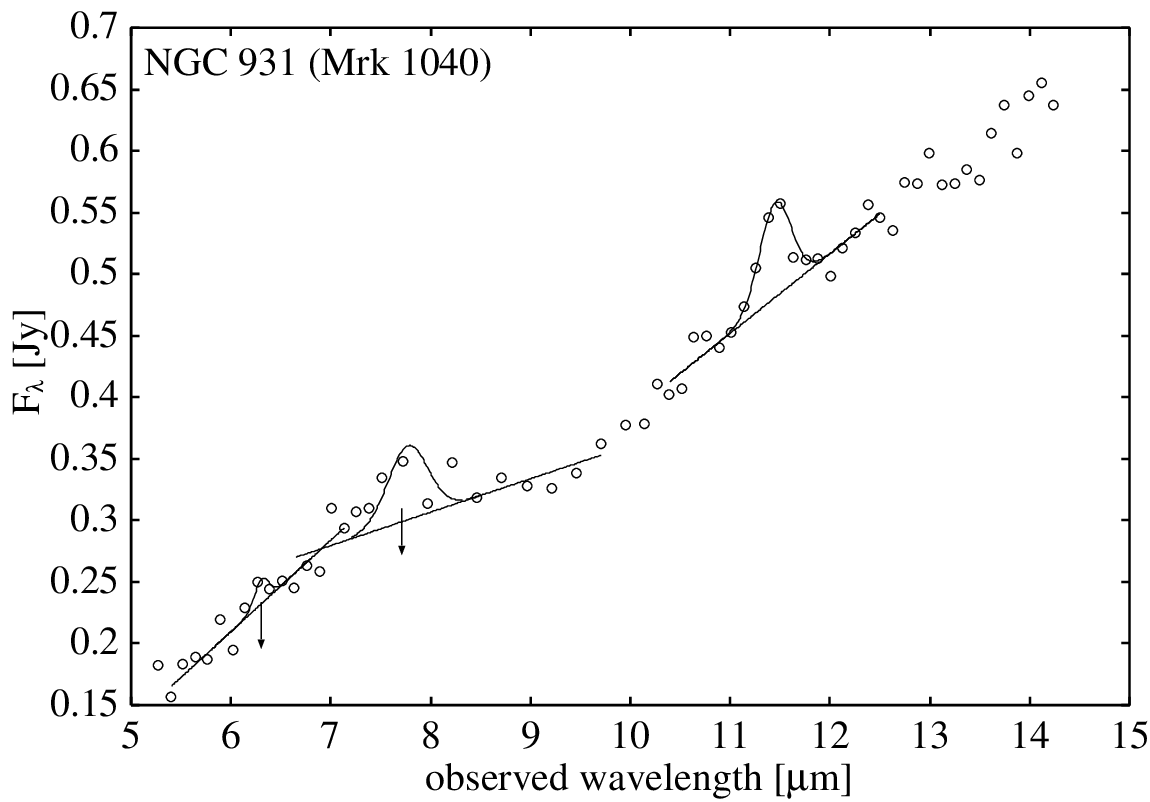}%931_1040
\plotone{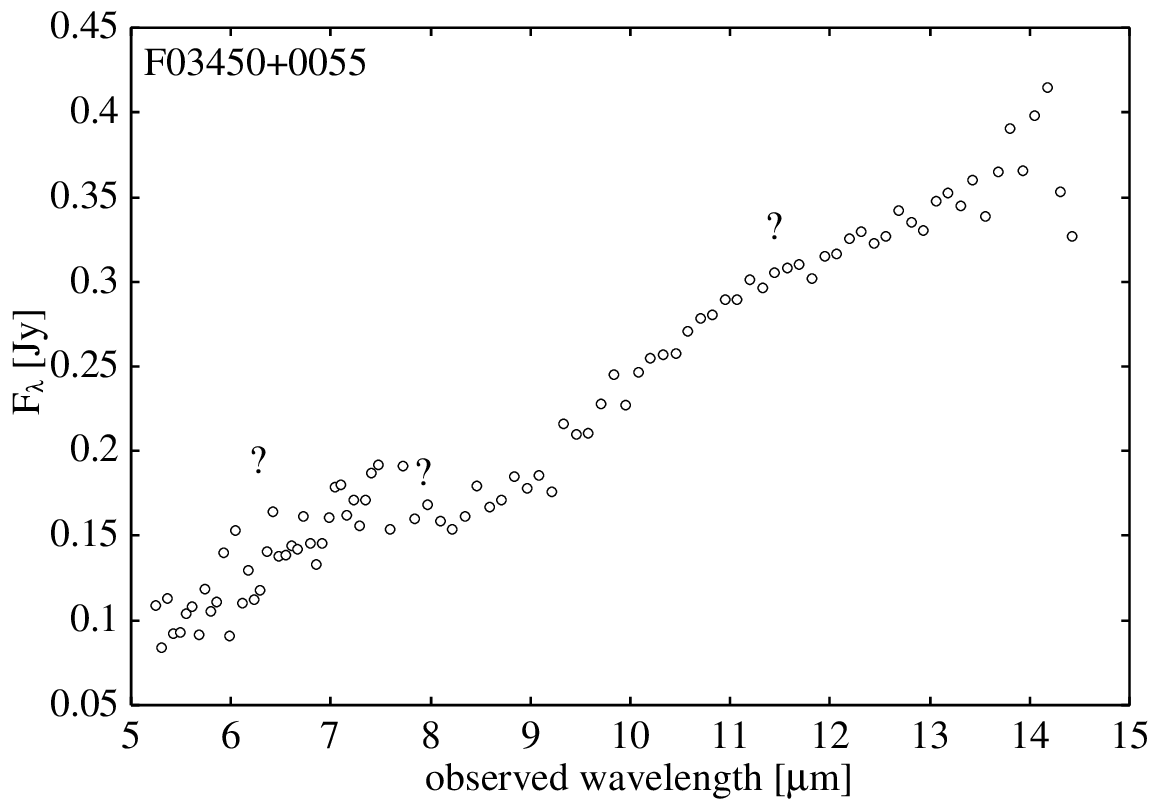}\\%03450
\plotone{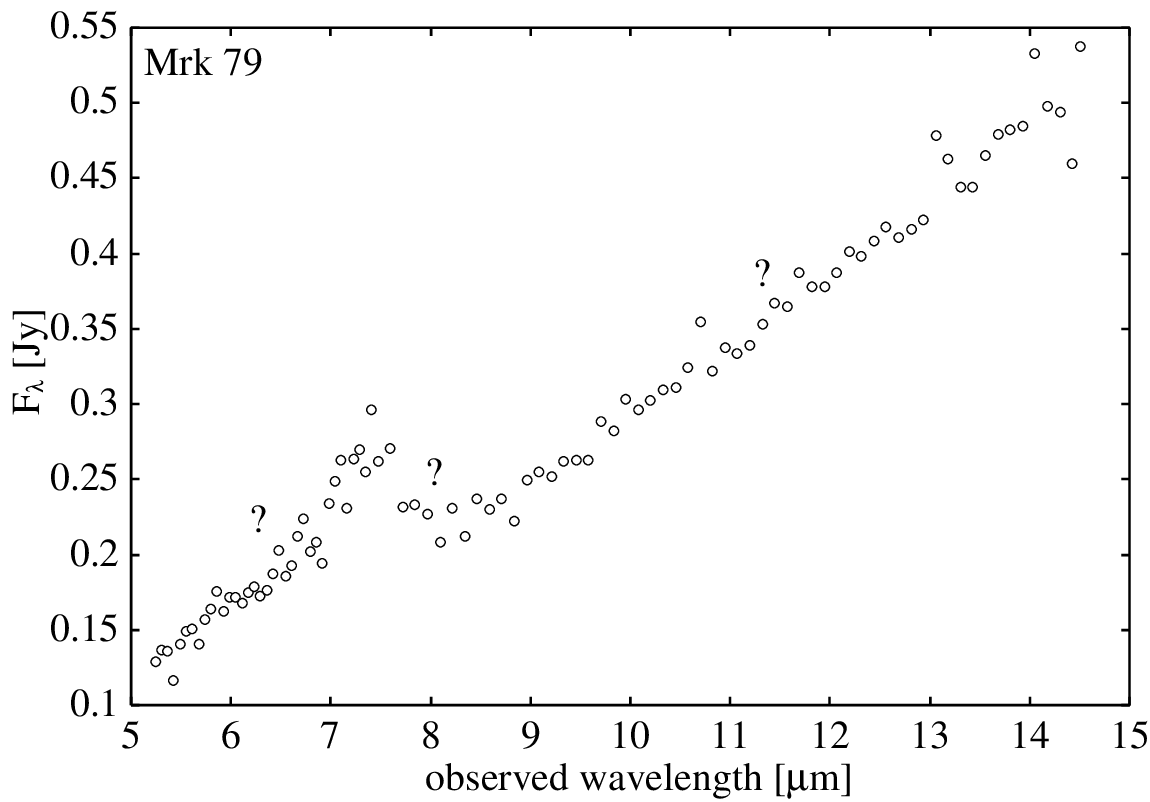}%79b2
\plotone{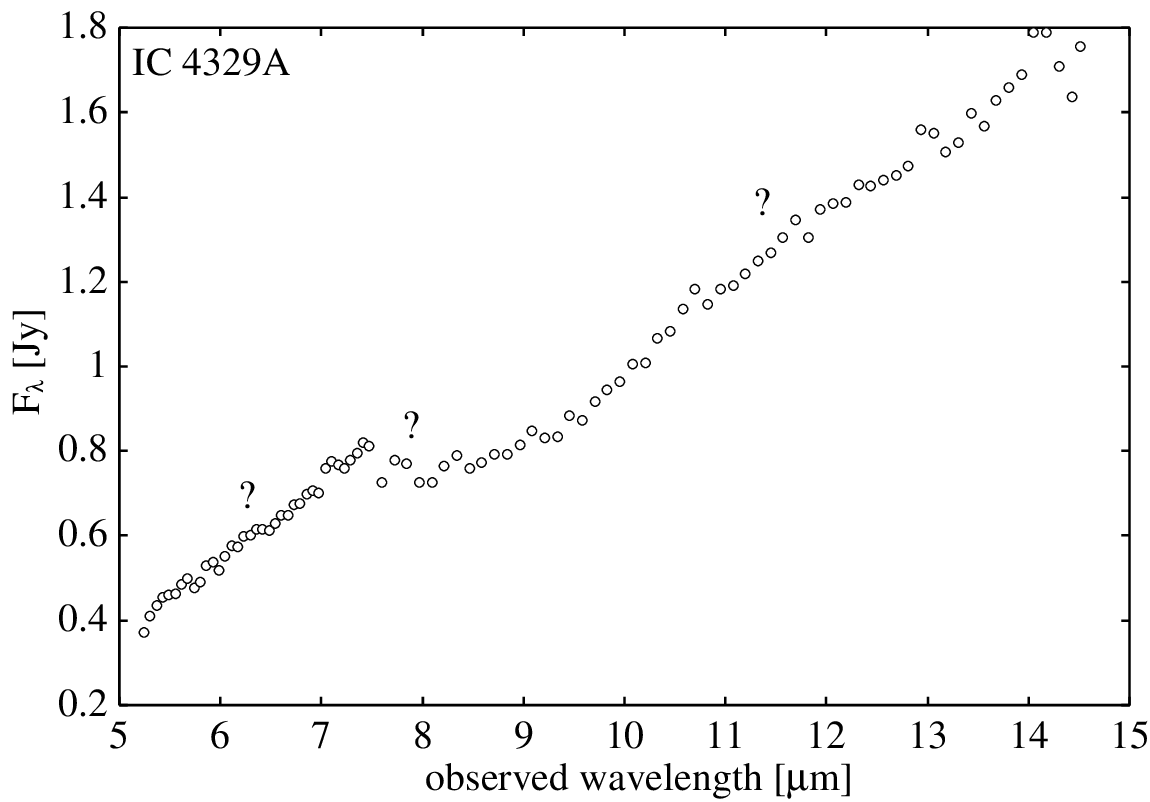}\\%4329b2
\caption{$5.2-14.5 \mu$m spectra of Seyfert 1 galaxies. The horizontal and vertical axes are the observed wavelength in $\mu$m and $F_{\lambda}$ in Jy, respectively. The solid straight lines are the adopted continuum to estimate each PAH emission fluxes. The solid curves are Gaussian profile fittings of PAH emission. The lower arrows and the question marks indicate the weak (lower than 3$\sigma$) and undetected PAH emission, respectively.}
\label{fig3}
\end{figure}

\begin{figure}
\epsscale{.40}
\plotone{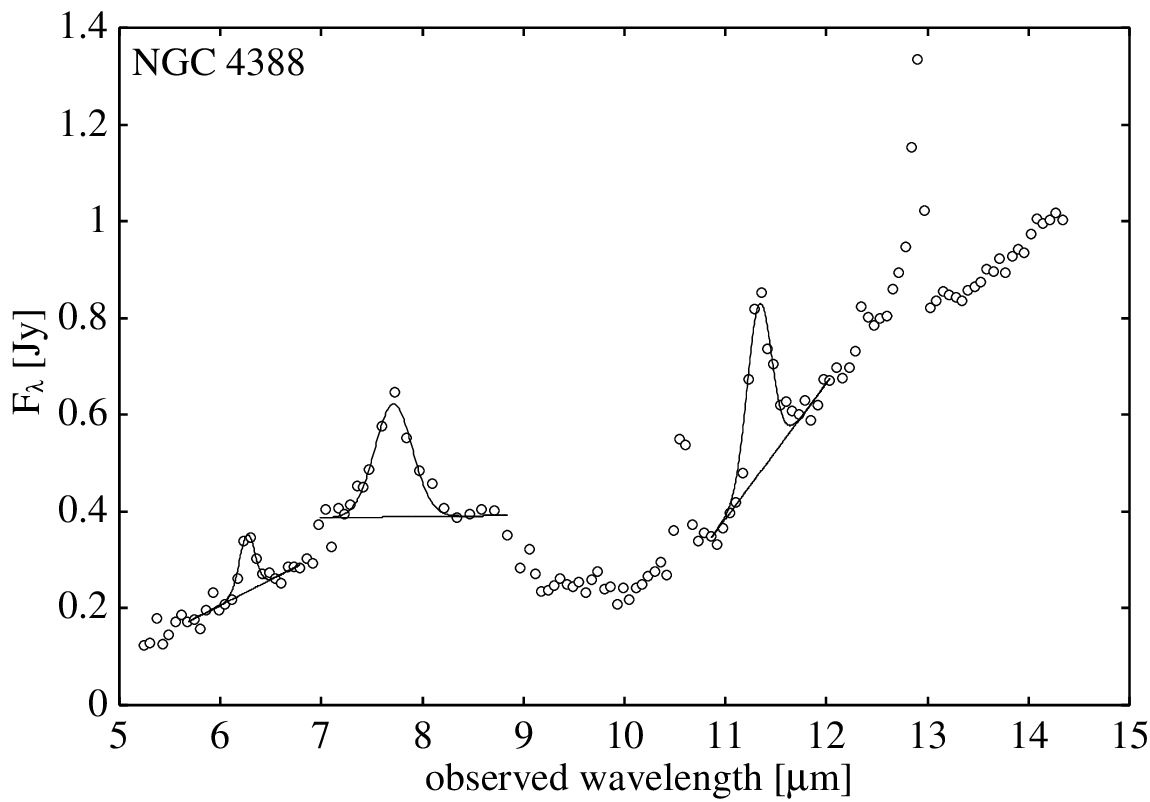}%4388
\plotone{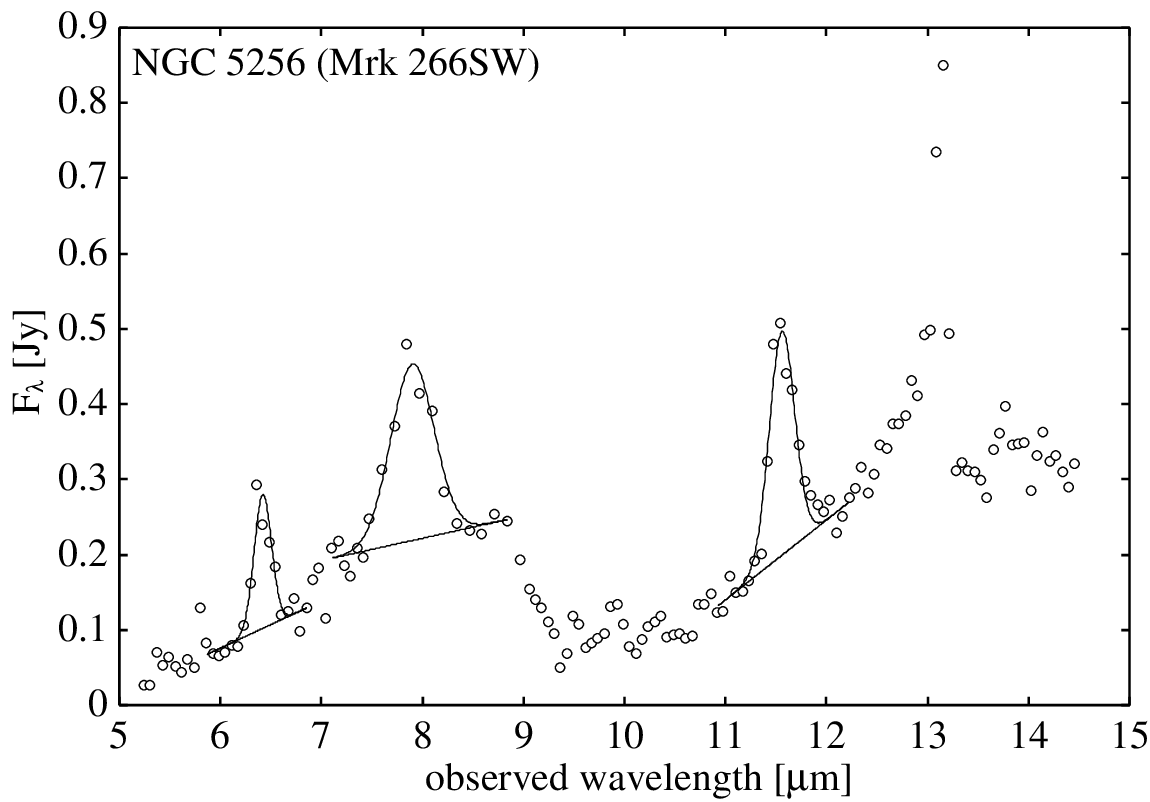}\\%5256_266
\plotone{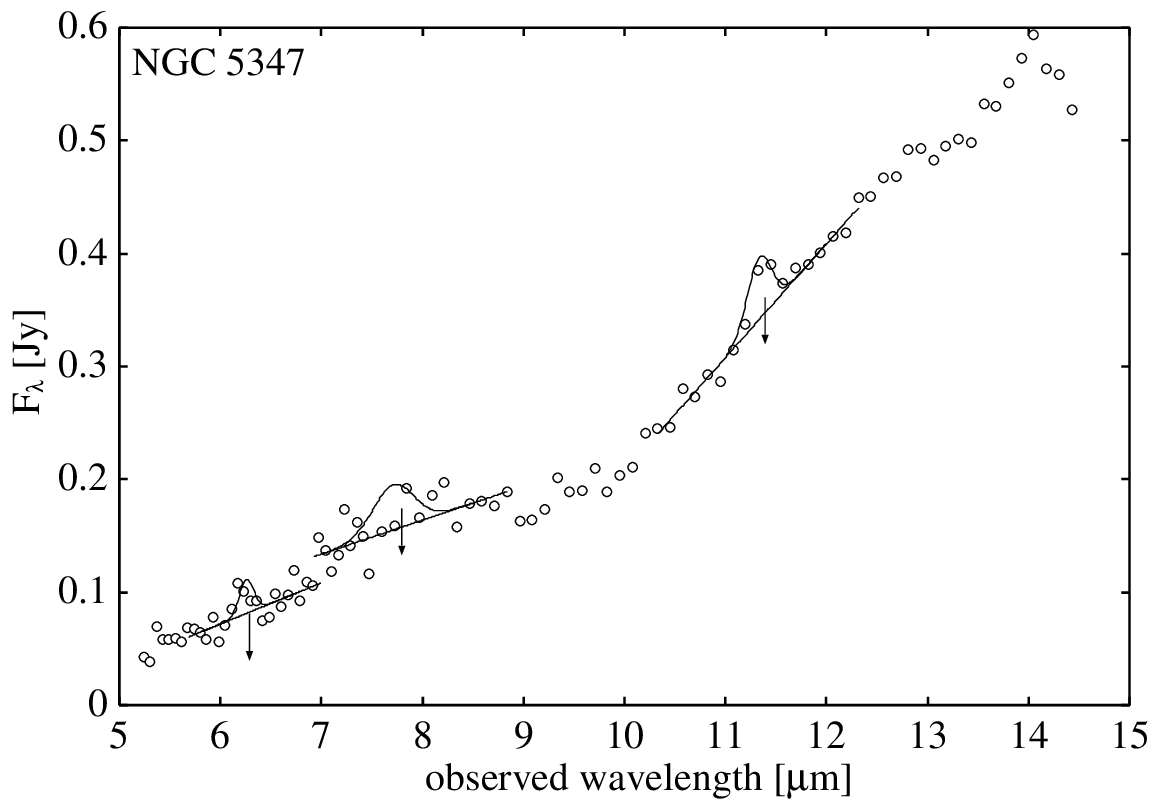}%5347
\plotone{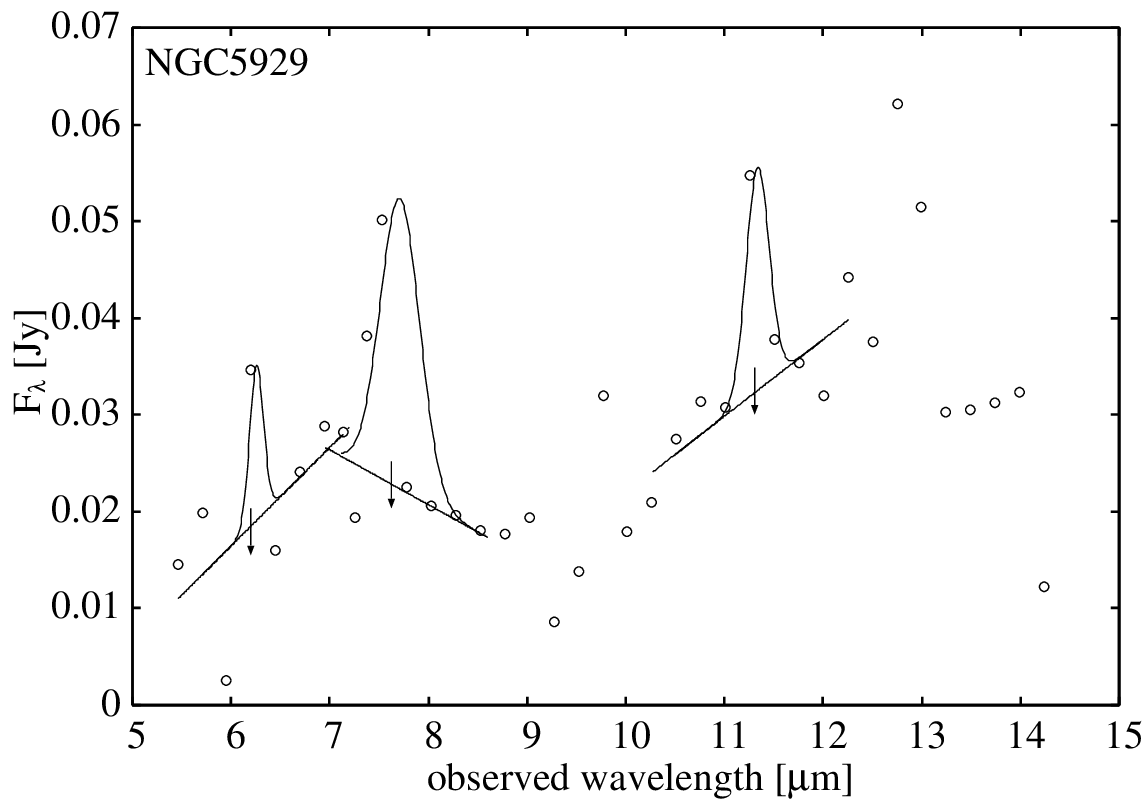}\\%5929
\plotone{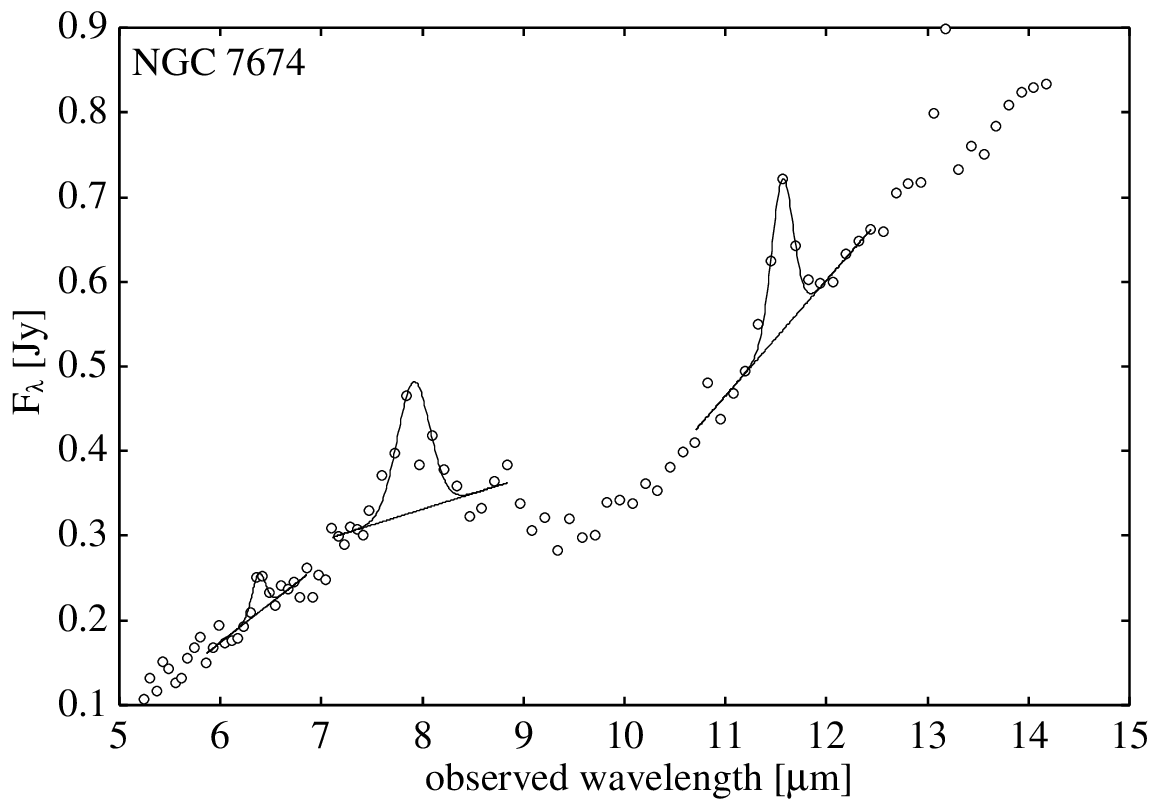}%7674
\plotone{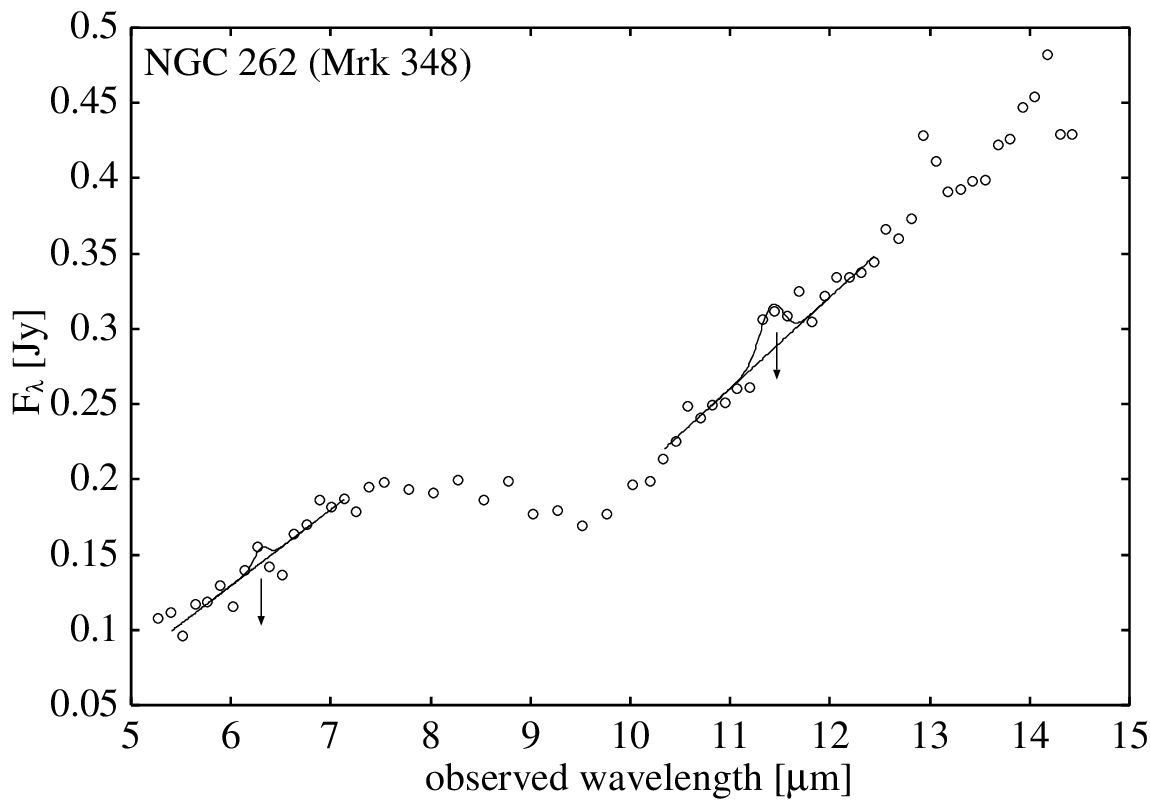}\\%262_348
\plotone{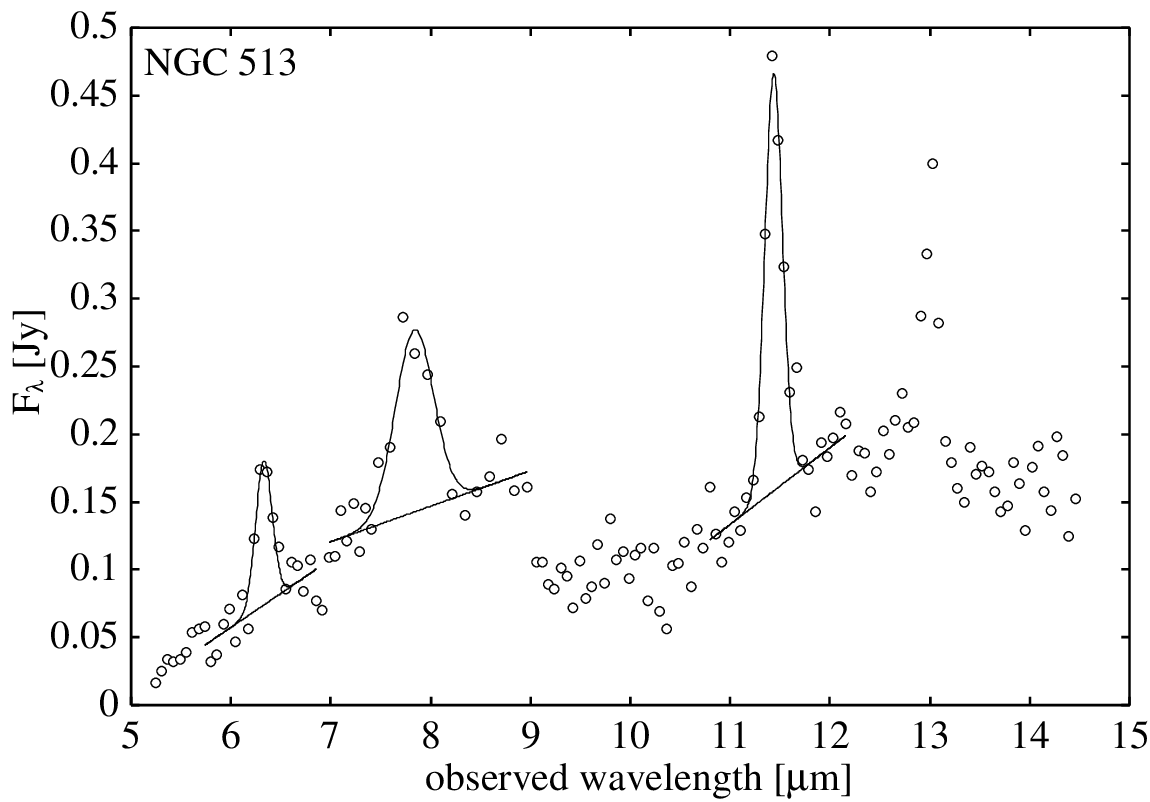}%513
\plotone{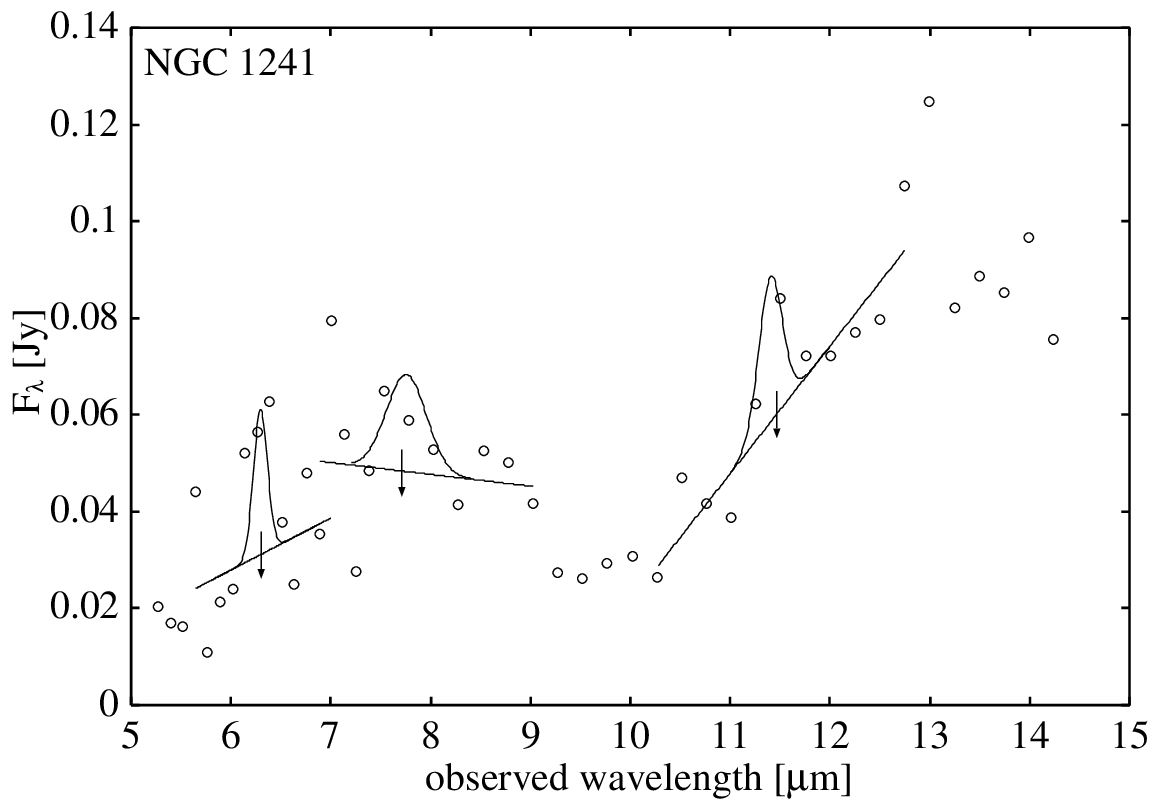}\\%1241
\plotone{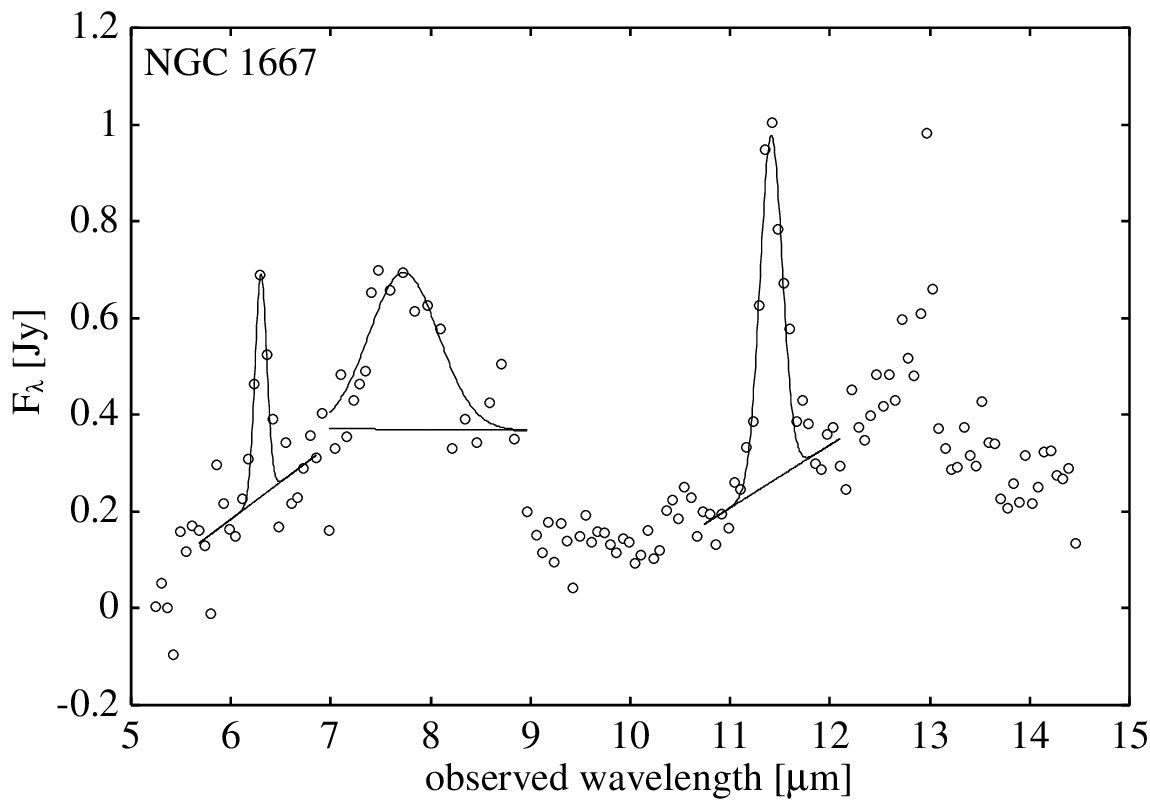}%1667
\caption{Same as Fig. \ref{fig3}, but these spectra are Seyfert 2 galaxies}
\label{fig4}
\end{figure}

Figure \ref{fig3} and \ref{fig4} represent 5.2-14.5 $\mu$m spectra of Seyfert 1 and 2 galaxies, respectively. Most of the Seyfert galaxies in Figure \ref{fig3} and \ref{fig4} show clearly detectable PAH emission features at $\lambda_{\rm rest} = 6.2 \mu \rm m$, $7.7 \mu \rm m$, and $11.3 \mu \rm m$. 

To estimate the fluxes of the PAH emission features at $\lambda_{\rm rest}$ 
= 6.2 $\mu$m, 7.7 $\mu$m, and 11.3 $\mu$m, we adopted a linear continuum 
determined from data points at the shorter and longer wavelength sides 
of individual PAH features, as used by \citet{Im07}. 
The adopted continuum levels are shown as solid straight lines in 
Fig. \ref{fig3} and \ref{fig4}. We fitted the PAH emission features with Gaussian profiles, which reproduced the observed data reasonably well. We treated the weak (lower than 3 $\sigma$) or undetected PAH emission as the upper limit or non-flux, respectively.

By using 3.3 $\mu$m, 6.2 $\mu$m, and 11.3 $\mu$m PAH luminosities, we can roughly estimate infrared luminosities of the nuclear starburst and total starburst. In Table \ref{tbl-4}, we showed infrared luminosities of the nuclear starburst estimated from the 3.3 $\mu$m PAH luminosity; $L_{\rm IR, 3.3} = L_{3.3}\times 10^{3}$ \citep{Mo90, Im02} and the total starburst estimated from the 6.2 $\mu$m and 11.3 $\mu$m PAH luminosity;  $L_{\rm IR, 6.2} = L_{6.2} \times 3 \times 10^2 $ \citep{Pe04} and $L_{\rm IR, 11.3} = L_{11.3} \times 7\times 10^2 $ \citep{So02}, respectively. Infrared luminosities of the total starbursts are several times or more as large as those of nuclear starbursts. This means that the circumnuclear starbursts dominate the total starburst luminosity. Thus, we can consider the 6.2 $\mu$m, 7.7 $\mu$m, and 11.3 $\mu$m PAH emission, which were observed in the whole host galaxy regions, as good indicators of a circumnuclear starburst.

\subsection{Luminosity Correlation between Nuclear/Circumnuclear Starburst and AGN}

\begin{figure}
\epsscale{.50}
\plotone{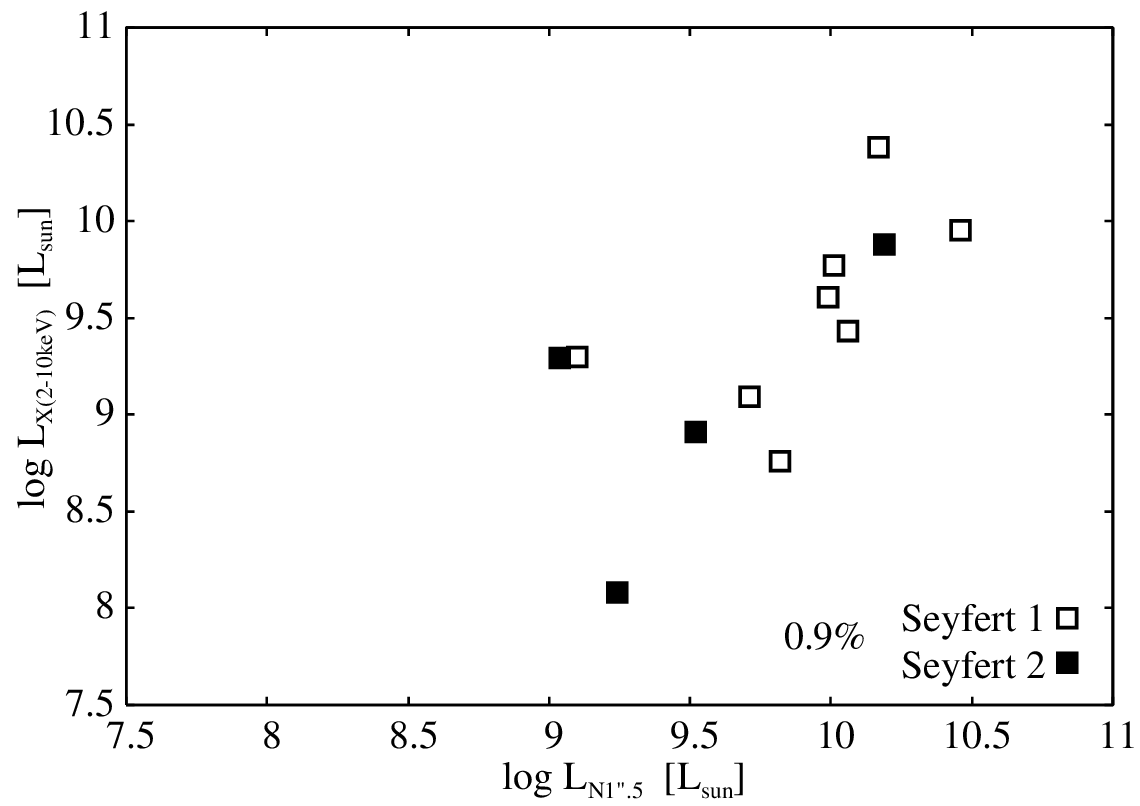}
\caption{The horizontal and the vertical axes are the nuclear N-band luminosity and the absorption-corrected hard (2-10 keV) X-ray luminosity, respectively. The percentage represents the probability that a correlation is not present.}
\label{fig5}
\end{figure}

In Figure \ref{fig5}, 
we checked a correlation between the nuclear N-band luminosity and the hard X-ray luminosity. 
We applied the generalized Kendall rank correlation statistics \citep{Is86} provided in the Astronomy Survival Analysis package \citep[ASURV;][]{FN85,Is86} to both types of Seyfert galaxies. The probability that a correlation is not present was 0.9 \% (2.6 $\sigma$). Since these hard X-ray luminosity are the absorption-corrected luminosity and estimated for the Compton-thin AGN, this correlation means that the nuclear N-band luminosity is good indicator of AGN activity for these Seyfert galaxies samples.

\begin{figure}
\epsscale{.50}
\plotone{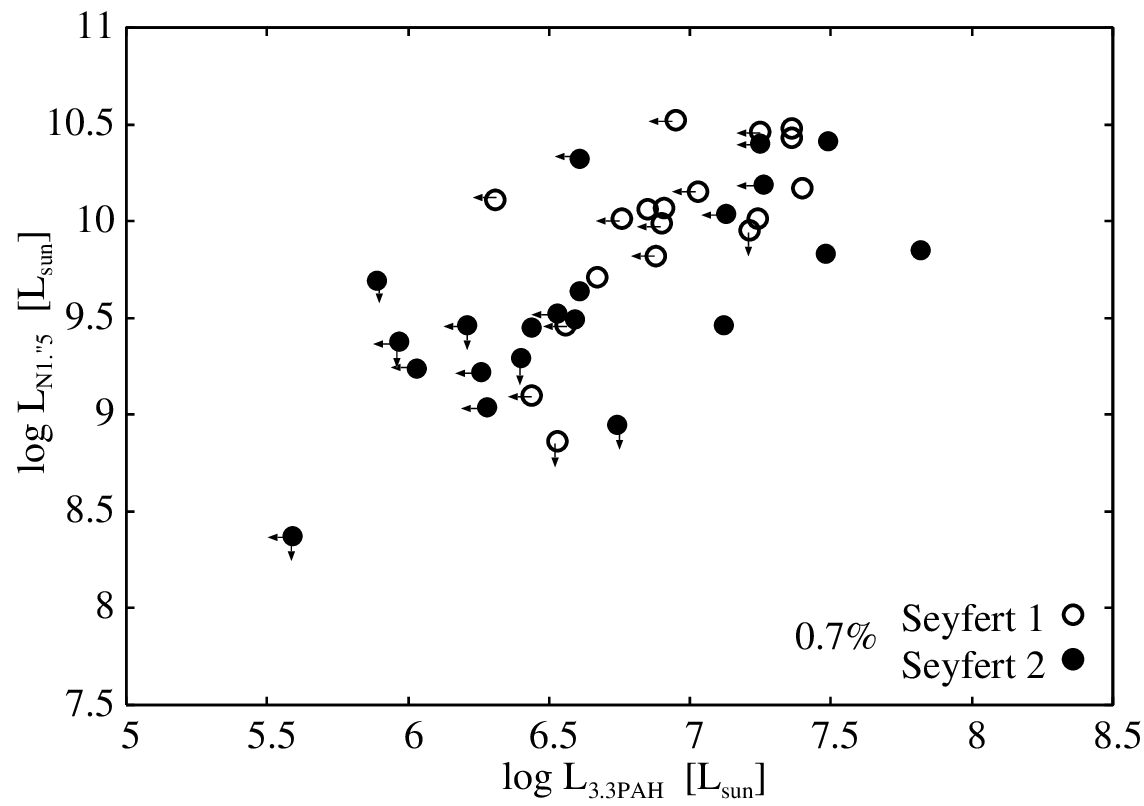}
\plotone{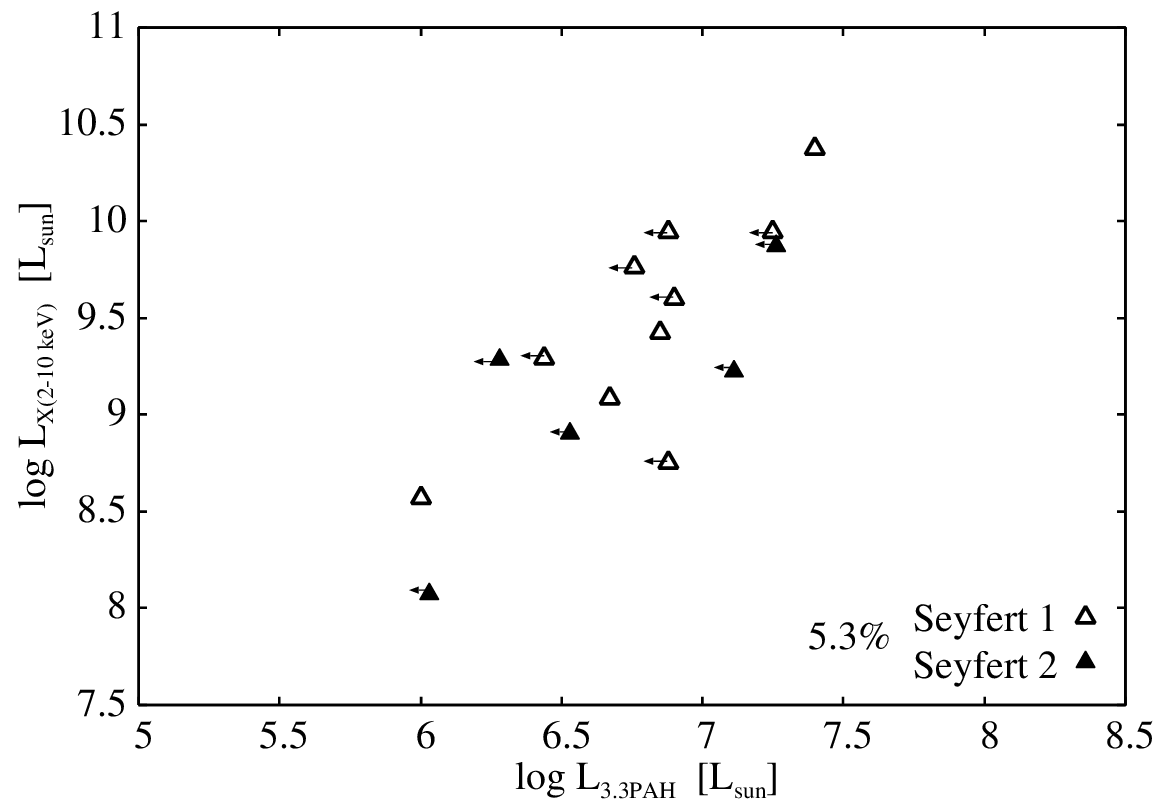}
\caption{{\it upper}: The horizontal axis is nuclear 3.3 $\mu \rm m$ PAH emission luminosity detected inside slit spectra (\citet{IW04} and this paper). The vertical axis represents the nuclear N-band (10.5 $\mu \rm m$) luminosity measured with ground-based two-dimensional camera with a 1".5 aperture \citep{Go04}. {\it bottom}: Same as upper figure, but the vertical axis represents the absorption-corrected hard (2-10 keV) X-ray luminosity. The percentage represents the probability that a correlation is not present.}
\label{fig6}
\end{figure}

In Figure \ref{fig6}, we plotted the nuclear N-band luminosity and the hard X-ray luminosity versus the 3.3 $\mu$m PAH emission luminosity. The probability that a correlation is not present was 0.7 \% (2.7 $\sigma$) and 5.3 \% (1.9 $\sigma$) for the nuclear N-band luminosity and the hard X-ray luminosity, respectively. \citet{IW04} shows the positive correlation between the 3.3 $\mu$m luminosity and the nuclear N-band luminosity. We also obtained the same results for the hard X-ray luminosity. Thus, the correlation between the luminosity of the nuclear starbursts and central AGN is also statistically confirmed in Seyfert galaxies in hard X-ray luminosity, which is directly involved in AGN activity. These nuclear N-band and hard X-ray results definitely show that nuclear starburst luminosity is positively correlated with AGN power.

For hard X-ray luminosity - 3.3 $\mu$m PAH luminosity correlation, the correlation probability here is lower than that of the nuclear N-band luminosity. This difference may result from a large number of upper/lower limit data and few sample data. To analyze hard X-ray data in detail, we may need to eliminate these uncertainties and collect more data. We therefore focus on only the nuclear N-band luminosity for the following results.

\begin{figure}
\epsscale{.50}
\plotone{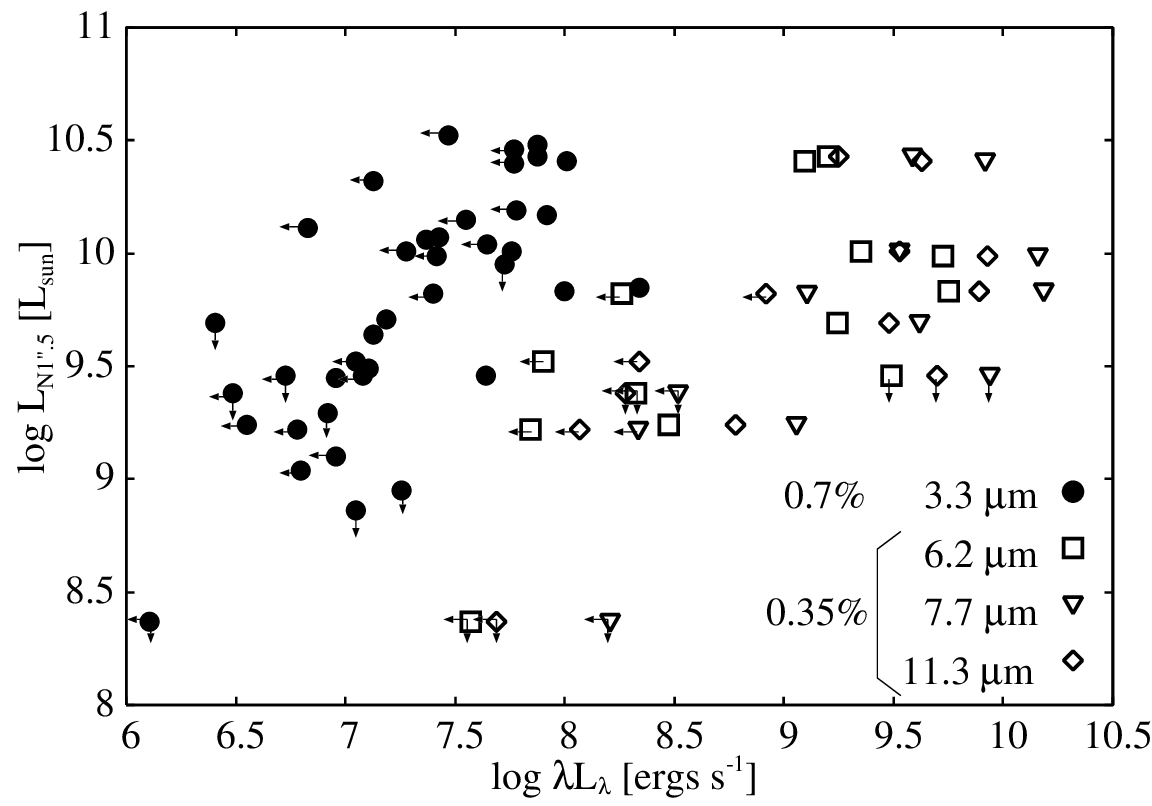}
\caption{The horizontal axis is the energy of the 3.3 $\mu$m, 6.2 $\mu$m, 7.7 $\mu$m, and 11.3 $\mu$m PAH emission and the vertical axis is the nuclear N-band luminosity. The percentages represent the probability that a correlation is not present.}
\label{fig7}
\end{figure}

In Figure \ref{fig7}, we plotted the nuclear N-band luminosity versus the energy for 6.2 $\mu$m, 7.7 $\mu$m, and 11.3 $\mu$m PAH emission. Here, we considered the 6.2 $\mu$m, 7.7 $\mu$m, and 11.3 $\mu$m PAH emission and plotted their energies. Since these PAH emission cover the whole host galaxy, we can consider them as indicators of circumnuclear starbursts.  We also plotted the energy for the 3.3 $\mu$m PAH emission in Fig. \ref{fig7} and applied the generalized Kendall rank correlation statistics to these 6.2 $\mu$m, 7.7 $\mu$m, and 11.3 $\mu$m PAH emission simultaneously. The probability that a correlation is not present was found to be 0.35 \% (2.9 $\sigma$) for these PAH emission. Thus, not only nuclear starburst luminosity but also circumnuclear starburst luminosity are connected to AGN luminosity.

\subsection{AGN Mass Accretion and Star-Forming Activity}

\begin{figure}
\epsscale{.50}
\plotone{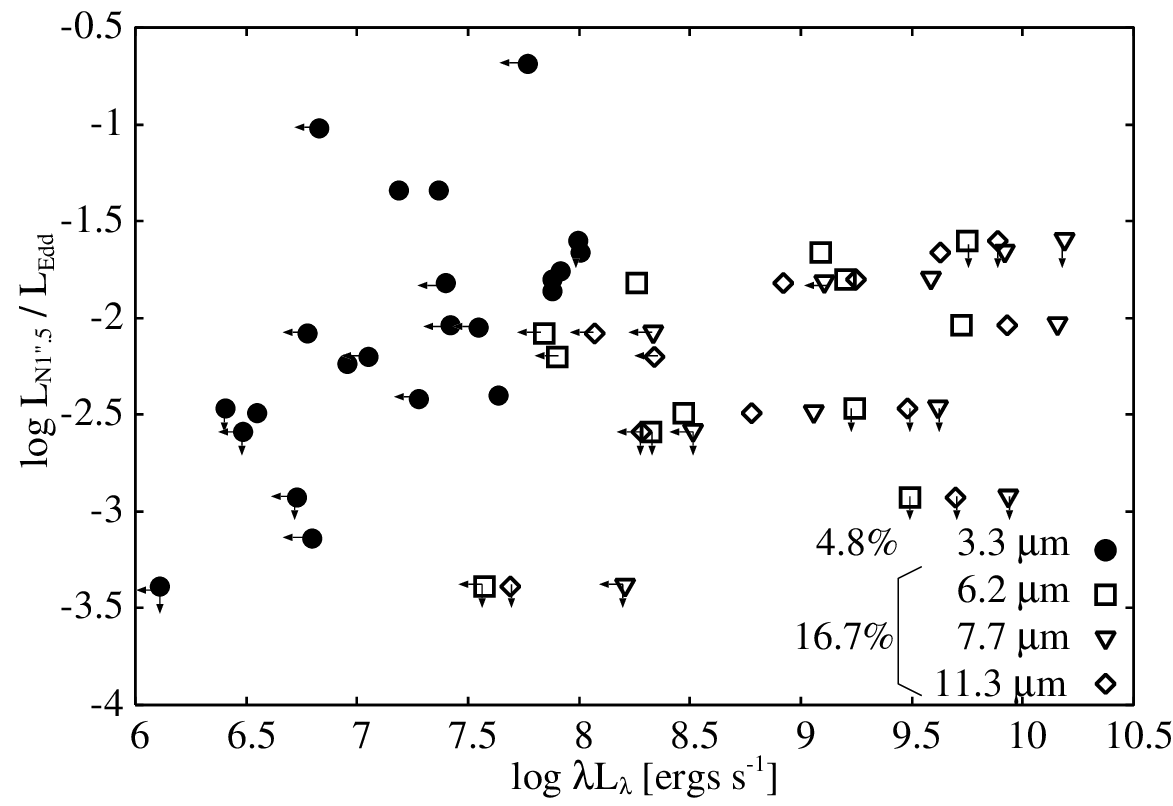}
\caption{Same as Fig. \ref{fig7}, but the vertical axis represents the ratio of the nuclear N-band luminosity to the AGN Eddington luminosity.}
\label{fig8}
\end{figure}

In Figure \ref{fig8}, we also plotted the ratio of the nuclear N-band luminosity to the AGN Eddington luminosity, $L_{\rm N1".5}/L_{\rm Edd}$ versus the energy of the nuclear/circumnuclear starbursts. The probability that a correlation is not present was found to be 4.8 \% (2.0 $\sigma$) for the nuclear starbursts (3.3 $\mu$m PAH emission) and 16.7 \% (1.4 $\sigma$)  for the circumnuclear starbursts (the other PAH emission). Thus, we statistically confirmed that the nuclear starburst is also positively correlated with the AGN luminosity normalized with the Eddington luminosity, whereas the circumnuclear starburst is only weakly correlated.

Here PAH emission are associated with star formation, and so the horizontal axis is also interpreted as an indicator of star-forming activity. The vertical axis also represents the mass accretion rate normalized with the Eddington mass accretion rate. This ratio corresponds to the efficiency of gas accretion onto a given SMBH mass. Thus, these results could be interpreted as the relationship between the efficiency of gas accretion onto a given SMBH mass and the star-forming activity around the AGN. Our results show a close correlation between the star-forming activity that is nearer the AGN and the efficiency of gas accretion. Therefore, starbursts near the AGN could affect AGN mass accretion more effectively.

%%%%%%%%%%%%
%%% Discussion %%%
%%%%%%%%%%%%

\section{Discussion: Link between AGN Mass Accretion and Nuclear/Circumnuclear Starbursts}

We found that nuclear and circumnuclear starburst luminosities are positively correlated with AGN luminosity. However, nuclear starburst luminosity compared to circumnuclear luminosity is correlated with AGN mass accretion more strongly. These results imply that the AGN mass accretion could be connected the star formation activity around AGN. Therefore, here, we discuss the interpretation of our results focusing on the effect of star formation activities.

\citet{Um97,Um98} showed the mass accretion onto the galactic center by the radiation effect from a circumnuclear starburst. They assumed a circumnuclear starburst ring and a gas disk within starburst ring. 

Through radial radiation pressure from the circumnuclear starburst, the optically thin surface layer of the gas disk (or the whole of the optically thin disk) is made to contract. Then the surface layer sheds angular momentum due to radiation drag. Also, since the radiation drag timescale for the dusty gas has been estimated a few $10^{6}$ yr and the radial radiation pressure can influence at shorter timescale further (Umemura et al. 1998), due to these mechanisms, luminous starbursts could carry a larger amount of gas toward the inner region within the duration of the starburst phase; $10^7$ yr \citep[e.g.,][]{Ef00} and the typical age for AGNs; $10^8$ yr (basically the Eddington timescale).

Radial radiation pressure may induce star formation as well as contraction of the gas disk shock \citep{Um98}. The superbubble driven by sequential explosions of supernovae in an OB association located in the plane-stratified gas distribution may also compress the surrounding gas due to shock \citep{SF76,TI86,NI89} and new stars may be born there. In fact, \citet{Ma05} showed the clear evidence of such a self-induced starburst at the inner edge of the expanding molecular superbubble in M82. This effect is even more expected in the case of a luminous starburst. If star formation is initiated in the contracted gas, it may induce further radiation effects and/or a superbubble in the inner gas disk. If the starbursts occur in such a sequential manner, the gas is carried farther into the inner region, and sequential starbursts may produce nuclear starbursts with some time lag (e.g., the timescale of the radiation pressure and radiation drag and/or the expansion time of the superbubble). 

Also, \citet{Fu00} showed that the gas ring in a barred galaxy could be unstable to gravitational instability and fragment into gas clumps. These clumps fall to the center by the torque from the massive clumps and/or by the dynamical friction from the stellar component \citep{Ma97}. Such dynamical effects could be also related to the gas accretion toward the center and produce nuclear starbursts.

Although the nuclear starburst is hidden by dusty gas, the structure is expected to be clumpy \citep{WN02}. In this case, not only the chaotic disturbance of the gas but also the radiation drag from the nuclear starburst may be able to reduce the angular momentum of the clumpy gas \citep{KU02}. Therefore, this innermost starburst could be directly related to the AGN mass accretion.

%%%%%%%%%%%%
%%% conclusion %%%
%%%%%%%%%%%%

\section{CONCLUSION}

To determine whether the starbursts influence AGN activities, we investigated the correlation between nuclear/circumnuclear starbursts and AGN activities. We found that circumnuclear starburst luminosity as well as the nuclear starburst luminosity is positively correlated with AGN luminosity. Moreover, nuclear starburst is also positively correlated with the AGN luminosity normalized with the AGN Eddington luminosity, whereas the circumnuclear starburst is only weakly correlated. This implies that close connections exist between starbursts nearer the AGN and AGN mass accretion.

\acknowledgments
We thank K. Wada and T. Nagao for providing valuable discussions. We also thank the anonymous referee for valuable comments. Y.W. is supported by the Research Fellowship of the JSPS for Young Scientists. M.I. is supported by Grants-in-Aid for Scientific Research (19740109).

%%%%%%%%%%%%
%%% Appendix %%%
%%%%%%%%%%%%

\section*{Appendix}
The infrared 3--4 $\mu$m spectra of the eight Seyfert galaxies were 
taken using IRTF SpeX \citep{Ra03}.  The 1.9--4.2 $\mu$m
cross-dispersed mode with a 0\farcs8 wide slit was employed.  This
mode enables $L$- (2.8--4.1 $\mu$m) and $K$-band (2--2.5 $\mu$m)
spectra to be obtained simultaneously, with a spectral resolution of R
$\sim$ 1000. The sky conditions were photometric throughout the
observations, and the seeing at $K$ was measured to be in the range
0$\farcs$35--0$\farcs$45 FWHM. A standard telescope nodding technique
(ABBA pattern) with a throw of 7$\farcs$5 was employed along the slit.
The telescope tracking was monitored with the infrared slit-viewer of
SpeX.  Each exposure was 15 sec, and 2 coadds were made at each
position.

F- or G-type main sequence stars (Table \ref{tbl-1}) were observed as
standard stars, with mean airmass difference of $<$0.1 to the individual
Seyfert nuclei, to correct for the transmission of the Earth's
atmosphere and to provide flux calibration.  The $L$-band magnitudes of
the standard stars were estimated from their $V$-band ($\lambda =$ 0.6
$\mu$m) magnitudes, adopting the $V-L$ colors appropriate to the stellar
types of individual standard stars \citep{To00}.

Standard data analysis procedures were employed, using IRAF   
\footnote{ IRAF is distributed by the National Optical Astronomy
Observatories, which are operated by the Association of Universities
for Research in Astronomy, Inc. (AURA), under cooperative agreement
with the National Science Foundation.}.
%---
Initially, frames taken with an A (or B) beam were subtracted from
frames subsequently taken with a B (or A) beam, and the resulting
subtracted frames were added and divided by a spectroscopic flat
image.  Then, bad pixels and pixels hit by cosmic rays were replaced
with the interpolated values of the surrounding pixels.  Finally the
spectra of Seyfert nuclei and standard stars were extracted, by
integrating signals over 1$\farcs$8--2$\farcs$2, depending on actual
signal profiles.  Wavelength calibration was performed taking into
account the wavelength-dependent transmission of the Earth's
atmosphere.  The spectra of Seyfert nuclei were divided by the
observed spectra of standard stars, multiplied by the spectra of
blackbodies with temperatures appropriate to individual standard stars
(Table \ref{tbl-1}).
Flux calibration was done based on signals of Seyfert galaxies and
standard stars detected inside our slit spectra. 
To reduce the scatter of data points, appropriate spectral binning is
employed, depending on the continuum flux levels of final spectra.

%%%%%%%%%%%%
%%% references %%%
%%%%%%%%%%%%

%%%%%%%%%
%%% table %%%
%%%%%%%%%

%--- Table 1 ---%
\begin{deluxetable}{llccclcccc}
%\rotate
\tabletypesize{\scriptsize}
\tablecaption{Observing log \label{tbl-1}}
\tablenum{1}
\tablewidth{0pt}
\tablehead{
\colhead{Object} & 
\colhead{Date} & 
\colhead{Telescope} & 
\colhead{Integration} & 
\colhead{P.A.}  & 
\multicolumn{4}{c}{Standard Stars} &
\colhead{Remark}  \\
\colhead{} & 
\colhead{(UT)} & 
\colhead{Instrument} & 
\colhead{(Min)} &
\colhead{($^{\circ}$)} &
\colhead{Name} &  
\colhead{$L$-mag} &  
\colhead{Type} &  
\colhead{$T_{\rm eff}$ (K)} &  
\colhead{} \\
\colhead{(1)} & \colhead{(2)} & \colhead{(3)} & \colhead{(4)} & 
\colhead{(5)} & \colhead{(6)} & \colhead{(7)} & \colhead{(8)} &
\colhead{(9)} & \colhead{(10)} 
}
\startdata
NGC 931 (Mrk 1040) & 2007 August 26 & IRTF SpeX & 96 & 0 & HR 720 & 4.4
& G0V & 5930 & 12$\mu$m Sy 1 \\
F03450+0055 & 2007 August 28 & IRTF SpeX & 112  & 0 & HR 962 & 3.7 & F8V
& 6000 & 12$\mu$m Sy 1 \\
NGC 262     & 2007 August 31 & IRTF SpeX & 80   & 0 & HR 410 & 5.0 & F7V
& 6240 & 12$\mu$m Sy 2 \\
NGC 513     & 2007 August 27 & IRTF SpeX & 120  & 0 & HR 410 & 5.0 & F7V
& 6240 & 12$\mu$m Sy 2 \\
MCG-2-8-39  & 2007 August 30 & IRTF SpeX & 136  & 0 & HR 784 & 4.5 & F6V
& 6400 & 12$\mu$m Sy 2 \\
MCG-3-58-7  & 2007 August 30 & IRTF SpeX & 80   & 0 & HR 8457 & 4.8 &
F6V & 6400 & 12$\mu$m Sy 2 \\
Mrk 993     & 2007 August 31 & IRTF SpeX & 112  & 0 & HR 410 & 5.0 &
F7V & 6240 & CfA Sy 2 \\
0152+06 (UGC 1395) & 2007 August 29 & IRTF SpeX & 120 & 0 & HR 508 & 4.7
& G3V & 5800 & CfA Sy 2 \\
\hline  
\enddata

\tablecomments{
Column (1): Object.
Col. (2): Observing date in UT.
Col. (3): Telescope and instrument. 
Col. (4): Net on-source integration time in min.
Col. (5): Position angle of the slit.
          0$^{\circ}$ corresponds to the north-south direction.
Col. (6): Standard star name.
Col. (7): Adopted $L$-band magnitude.
Col. (8): Stellar spectral type.
Col. (9): Effective temperature.
Col. (10): Seyfert 1 or 2 galaxy in the 12 $\mu$m or CfA sample.
}
\end{deluxetable}

\clearpage

%--- Table 2 ---%
\begin{deluxetable}{lccccc}
\tabletypesize{\scriptsize}
%\rotate
\tablecaption{Properties of the Nuclear 3.3 $\mu$m PAH Emission Feature
\label{tbl-2}} 
\tablenum{2}
\tablewidth{0pt}
\tablehead{
\colhead{Object} & 
\colhead{$f_{\rm nuclear-3.3 PAH}$} &
\colhead{$L_{\rm nuclear-3.3 PAH}$} & 
\colhead{rest EW$_{3.3 \rm PAH}$} \\ 
\colhead{} & 
\colhead{($\times$ 10$^{-14}$ ergs s$^{-1}$ cm$^{-2}$)} &
\colhead{($\times$ 10$^{40}$ ergs s$^{-1}$)} & 
\colhead{(nm)} \\
\colhead{(1)} & 
\colhead{(2)} & 
\colhead{(3)} & 
\colhead{(4)}
}
\startdata
NGC 931     & $<$6.5 & $<$2.9 & $<$2.5 \\
F03450+0055 & 1.8    & 3.1    & 1.0 \\
NGC 262     & $<$3.3 & $<$1.3 & $<$1.6 \\
NGC 513     & 0.5    & 0.3    & 3.5 \\
MCG-2-8-39  & $<$1.3 & $<$2.0 & $<$13 \\
MCG-3-58-7  & 1.8    & 3.3    & 1.1 \\
Mrk 993     & $<$2.2 & $<$0.9 & $<$16 \\
0152+06     & $<$1.6 & $<$0.8 & $<$13 \\
\enddata

\tablecomments{Column (1): Object. 
Col. (2): Observed nuclear 3.3 $\mu$m PAH flux.
Col. (3): Observed nuclear 3.3 $\mu$m PAH luminosity.
Col. (4): Rest frame equivalent width of the 3.3 $\mu$m PAH emission.
}
\end{deluxetable}

%--- Table 3 ---%
\begin{deluxetable}{lcccccccccccc}
\tabletypesize{\tiny}
%\rotate
\tablecaption{Seyfert galaxies data
\label{tbl-3}} 
\tablenum{3}
\tablewidth{0pt}
\tablehead{
\colhead{Object} & 
\colhead{$z$} &
\colhead{$\log L_{\rm 3.3} $} & 
\colhead{$\log L_{\rm 6.2} $} & 
\colhead{$\log L_{\rm 7.7} $} & 
\colhead{$\log L_{\rm 11.3} $} & 
\colhead{$\log L_{\rm N 1".5} $} & 
\colhead{$ \log M_{\rm BH} $} & 
\colhead{$\log \rm{E_{N}} $ } & 
\colhead{$\log L_{\rm X} $} & 
\colhead{references} \\ 
\colhead{} & 
\colhead{} &
\colhead{$(L_{\odot})$} & 
\colhead{$(L_{\odot})$} & 
\colhead{$(L_{\odot})$} & 
\colhead{$(L_{\odot})$} & 
\colhead{$(L_{\odot})$} & 
\colhead{$(M_{\odot})$} & 
\colhead{} & 
\colhead{$(L_{\odot})$} & 
\colhead{} \\
\colhead{(1)} & 
\colhead{(2)} & 
\colhead{(3)} & 
\colhead{(4)} &
\colhead{(5)} & 
\colhead{(6)} & 
\colhead{(7)} & 
\colhead{(8)} & 
\colhead{(9)} & 
\colhead{(10)} & 
\colhead{(11)} 
}
\startdata
Seyfert 1 (CfA Sample)&&&&&&&\\
\tableline
Mrk 335 & 0.025 & 6.85 & ? & ? & ? & 10.06 & 6.89 & $-1.34$ & 9.43 & 1\\
NGC 863 (Mrk 590) & 0.027 & $< $6.88 &---&---&---&---& 7.58 &---& 9.95 & 2\\
NGC 3786 (Mrk 744) & 0.009 & $< $6.36 &---&---&---&---& 7.53 &---&---&---\\
NGC 4235 & 0.008 & 6.00 &---&---&---&---&---&---& 8.58 & 3 \\
NGC 4253 (Mrk 766) & 0.013 & 6.67 &---&---&---& 9.71 & 6.54 & $-1.34$& 9.09 & 1 \\
NGC 5548 & 0.017 & $< $6.76 &---&---&---& 10.01 & 7.92 & $-2.42$& 9.77 & 1 \\
Mrk 817 & 0.031 & 7.36 & 8.41 & 8.7 & 8.2 & 10.43 & 7.72 & $-1.80$&---&---\\
NGC 7469 & 0.016 & $< $6.9 & 8.93 & 9.27 & 8.88 & 9.99 & 7.52 & $-2.04$&9.61&1 \\
Mrk 530 (NGC 7603) & 0.029 & 7.24 & 8.56 & 8.64 & 8.48 & 10.01 &---&---&---&---\\
\tableline
Seyfert 1 (12 $\rm \mu m$ Sample)&&&&&&&\\
\tableline
NGC 931 (Mrk 1040) & 0.016 & $< $6.88$^{\dagger}$ & $< $7.47 & $< $8.22 & 7.87 & 9.82 & 7.13 & $-1.82$&8.76& 1\\
F03450+0055 & 0.031 & 6.91$^{\dagger}$ & ? & ? & ? & 10.07 &---&---&---&---\\
3C 120 & 0.033 & 7.4 &---&---&---& 10.17 & 7.41 & $-1.76$&10.38&2\\
Mrk 618 & 0.035 & 7.21 &---&---&---& $< $9.95 &---&---&---&---\\
MCG -5-13-17 & 0.013 & $< $6.56 &---&---&---& 9.46 &---&---&---&---\\
Mrk 79 & 0.022 & $< $7.03 & ? & ? & ? & 10.15 & 7.68 & $-2.05$&---&---\\
Mrk 704 & 0.030 & $< $6.95 &---&---&---& 10.52 &---&---&---&---\\
NGC 2992 & 0.008 & $< $6.44 &---&---&---& 9.10 &---&---&9.30&2\\
Mrk 1239 & 0.019 & $< $6.31 &---&---&---& 10.11 & 6.62 & $-1.02$&---&---\\
MCG -2-33-34 & 0.014 & 6.53 &---&---&---& $< $8.86 &---&---&---&---\\
IC 4329A & 0.016 & $< $7.25 & ? & ? & ? & 10.46 & 6.64 & $-0.69$&9.95&1\\
Mrk 509 & 0.036 & 7.36 &---&---&---& 10.48 & 7.83 & $-1.86$&---&---\\
\tableline
\tableline
Seyfert 2 (CfA Sample)&&&&&&&\\
\tableline
NGC 4388 & 0.008 & $< $6.03 & 7.68 & 8.17 & 7.73 & 9.24 & 7.22 & $-2.49$&8.08&3\\
NGC 5252 & 0.023 & $< $7.11 &---&---&---&---&---&---&9.23&4\\
NGC 5256 (Mrk 266SW) & 0.028 & 7.48 & 8.96 & 9.30 & 8.84 & 9.83 & $> $6.92 & $< -1.60$ &---&---\\
NGC 5347 & 0.008 & $< $6.26 & $< $7.05 & $< $7.45 & $< $7.02 & 9.22 & 6.79 & $-2.08$&---&---\\
NGC 5929 & 0.008 & $< $5.59 & $< $6.78 & $< $7.32 & $< $6.64 & $< $8.37 & 7.25 & $< -3.39$&---&---\\
NGC 7674 & 0.029 & 7.49 & 8.30 & 9.03 & 8.58 & 10.41 & 7.56 & $-1.66$&---&---\\
\tableline
Seyfert 2 (12 $\rm \mu m$ Sample) &&&&&&&\\
\tableline
Mrk 938 & 0.019 & 7.82 &---&---&---& 9.85 &---&---&---&---\\
NGC 262 (Mrk 348) & 0.015 & $< $6.53$^{\dagger}$ & $< $7.11 & ? & $< $7.29 & 9.52 & 7.21 & $-2.20$&8.91&5\\
NGC 513 & 0.020 & 5.89$^{\dagger}$ & 8.45 & 8.73 & 8.43 & $< $9.69 & 7.65 & $< -2.47$&---&---\\
NGC 1125 & 0.011 & 6.74 &---&---&---& $< $8.95 &---&---&---&---\\
NGC 1241 & 0.014 & $< $5.97 & $< $7.53 & $< $7.63 & $< $7.23 & $< $9.38 & 7.46 & $< -2.59$&---&---\\
NGC 1320 (Mrk 607) & 0.010 & 6.44 &---&---&---& 9.45 & 7.18 & $-2.24$&---&---\\
F04385-0820 & 0.015 & 6.61 &---&---&---& 9.64 &---&---&---&---\\
NGC 1667 & 0.015 & $< $6.21 & 8.7 & 9.05 & 8.65 & $< $9.46 & 7.88 & $< -2.93$&---&---\\
NGC 3660 & 0.012 & 6.4 &---&---&---& $< $9.29 &---&---&---&---\\
NGC 4968 & 0.010 & 6.59 &---&---&---& 9.49 &---&---&---&---\\
MCG -3-34-64 & 0.017 & $< $6.61 &---&---&---& 10.32 &---&---&---&---\\
NGC 5135 & 0.014 & 7.12 &---&---&---& 9.46 & 7.35 & $-2.40$&---&---\\
MCG -2-40-4 (NGC 5995) & 0.024 & $< $7.26 &---&---&---&10.19 &---&---&9.88&5\\
F15480-0344 & 0.030 & $< $7.13 &---&---&---& 10.04 &---&---&---&---\\
NGC 7172 & 0.009 & $< $6.28 &---&---&---& 9.04 & 7.67 & $-3.14$&9.29&2\\
MCG -3-58-7 & 0.032 & $< $7.25 &---&---&---& 10.40 &---&---&---&---\\
\enddata

\tablecomments{Column (1): Object.
Col. (2): Redshift. 
Col. (3): PAH luminosity at 3.3 $\mu$m. $^{\dagger}$ Taken from this paper.
Col. (7): $L_{6.2}, L_{7.7}, L_{11.3}$, and $L_{N1".5} $ are the PAH luminosity at 6.2 $\mu$m, 7.7 $\mu$m, 11.2 $\mu$m, and the nuclear N-band $(10.5 \mu \rm{m})$, respectively. 
Col. (8): Black hole mass. 
Col. (9): Ratio of nuclear N-band luminosity to AGN Eddington luminosity. 
Col. (10) and (11): Absorption-corrected hard (2-10 keV) X-ray luminosity and corresponding references; 1, \citet{On05}; 2, \citet{Sh06}; 3, \citet{Pa06}; 4, \citet{Tu97}; 5, \citet{PB02}. "---" shows no information. "?" means undetected PAH emission.
}
\end{deluxetable}

%--- Table 4 ---%
\begin{deluxetable}{lccccccc}
\tabletypesize{\scriptsize}
%_rotate
\tablecaption{Infrared luminosity
\label{tbl-4}} 
\tablenum{4}
\tablewidth{0pt}
\tablehead{
\colhead{Object} & 
\colhead{$\log{L_{\rm IR, 3.3}}$} & 
\colhead{$\log{L_{\rm IR, 6.2}}$} & 
\colhead{$\log{L_{\rm IR, 11.3}}$} & 
\colhead{$L_{\rm IR, 6.2} / L_{\rm IR, 3.3}$} &
\colhead{$L_{\rm IR, 11.3} / L_{\rm IR, 3.3}$} \\
\colhead{} & 
\colhead{($L_{\odot}$)} &
\colhead{($L_{\odot}$)} &
\colhead{($L_{\odot}$)} &
\colhead{} &
\colhead{} \\
\colhead{(1)} & 
\colhead{(2)} & 
\colhead{(3)} & 
\colhead{(4)} &
\colhead{(5)} &
\colhead{(6)}
}
\startdata
Mrk 817              & 10.36 & 10.88 & 11.05 & 3.30   & 4.94\\
NGC 7469             & $<9.90$ & 11.40 & 11.73 & $>31.52$ & $>68.21$\\
Mrk 530 (NGC 7603)   & 10.24 & 11.03 & 11.33 & 6.14   & 12.41\\
NGC 931 (Mrk 1040)   & $<9.88$ & $<9.94$ & 10.72 & ---  & $>6.98$\\
NGC 4388             & $<9.03$ & 10.15 & 10.58 & $>13.14$ & $>35.80$\\
NGC 5256 (Mrk 266SW) & 10.48 & 11.43 & 11.69 & 8.88   & 16.36\\
NGC 5347  &$<9.26$  & $<9.52$  & $<9.87$  &---&---\\
NGC 5929  & $<8.59$ & $<9.25$ & $<9.49$ &---&---\\
NGC 7674             & 10.49 & 10.77 & 11.43 & 1.90   & 8.79\\
NGC 262 (Mrk 348)  & $<9.53$ & $<9.58$ & $<10.14$ &---&---\\
NGC 513              & 8.89  & 10.92 & 11.28 & 106.79 & 247.67\\
NGC 1241            & $<8.97$ & $<10.00$ & $<10.08$ &---&---\\
NGC 1667             & $<9.21$ & 11.17 & 11.50 & $>90.89$ & $>196.73$\\
\enddata

\tablecomments{Column (1): Object.
Column (2): Infrared luminosity of the nuclear starburst. $L_{\rm IR, 3.3} = L_{3.3} \times 10^{3}$ \citep{Mo90, Im02}.
Col. (3) and (4): $L_{\rm IR, 6.2} = L_{6.2} \times 3\times 10^{2}) $ \citep{Pe04} and $L_{\rm IR, 11.3} = L_{11.3} \times 7\times 10^2 $ \citep{So02} are the infrared luminosity of total starbursts, which were estimated from 6.2 $\mu$m and 11.3 $\mu$m PAH luminosity, respectivery.
Col. (5) and (6): Infrared luminosity Ratios of total starburst to nuclear starburst. 
"---" shows that both the infrared luminosity of the nuclear starburst and the total starburst are upper limit data.}
\end{deluxetable}

%%%%%%%%%%
%%% figure %%%
%%%%%%%%%%

%---  Figure 1 ---%
\clearpage

\clearpage

%---  Figure 2 ---%
\clearpage

%---  Figure 3 ---%

%---  Figure 4 ---%

%---  Figure 5 ---%

%---  Figure 6 ---%

%---  Figure 7 ---%

%---  Figure 8 ---%


\begin{thebibliography}{}
\bibitem[Barnes \& Hernquist(1991)]{BH91} Barnes, J. E., \& 
Hernquist, L. E.\ 1991, \apjl, 370, L65
\bibitem[Barnes \& Hernquist(1992)]{BH92} Barnes, J. E., \& 
Hernquist, L.\ 1992, \araa, 30, 705 
\bibitem[Barth et al.(1995)]{Ba95} Barth, A. J., Ho, L. C., 
Filippenko, A. V., \& Sargent, W. L.\ 1995, \aj, 110, 1009
\bibitem[Benedict et al.(1996)]{Be96} Benedict, G. F., 
Smith, B. J., \& Kenney, J. D. P.\ 1996, \aj, 112, 1318 
\bibitem[Bian \& Gu(2007)]{BG07} Bian, W., \& Gu, Q.\ 2007, 
\apj, 657, 159 
\bibitem[Buta et al.(1995)]{Bu95} Buta, R., Purcell, G. B., 
\& Crocker, D. A.\ 1995, \aj, 110, 1588 
\bibitem[Davies et al.(2007)]{Da07} Davies, R., Mueller 
Sanchez, F., Genzel, R., Tacconi, L., Hicks, E, Friedrich, S, \& Sternberg, 
A.\ 2007, ArXiv e-prints, 704, arXiv:0704.1374 
\bibitem[Efstathiou et al.(2000)]{Ef00} Efstathiou, A., 
Rowan-Robinson, M., \& Siebenmorgen, R.\ 2000, \mnras, 313, 734
\bibitem[Elmouttie et al.(1998)]{El98} Elmouttie, M., 
Koribalski, B., Gordon, S., Taylor, K., Houghton, S., Lavezzi, T., Haynes, 
R., \& Jones, K.\ 1998, \mnras, 297, 49 
\bibitem[Feigelson \& Nelson(1985)]{FN85} Feigelson, E. D., \& Nelson, P. I.\ 1985, \apj, 293, 192
\bibitem[Forbes et al.(1994)]{Fo94} Forbes, D. A., Norris, 
R. P., Williger, G. M., \& Smith, R. C.\ 1994, \aj, 107, 984 
\bibitem[Fukuda et al.(2000)]{Fu00} Fukuda, H., Habe, A., \& 
Wada, K.\ 2000, \apj, 529, 109 
\bibitem[Fukuda et al.(1998)]{Fu98} Fukuda, H., Wada, K., \& 
Habe, A.\ 1998, \mnras, 295, 463 
\bibitem[Fukue et al.(1997)]{Fu97} Fukue, J., Umemura, M., 
\& Mineshige, S.\ 1997, \pasj, 49, 673
\bibitem[Gorjian et al.(2004)]{Go04} Gorjian, V., Werner, 
M. W., Jarrett, T. H., Cole, D. M., \& Ressler, M. E.\ 2004, \apj, 605, 156 
\bibitem[Hernquist(1989)]{He89} Hernquist, L.\ 1989, \nat, 
340, 687
\bibitem[Houck et al.(2004)]{Ho04} Houck, J. R., et al.\ 
2004, \apjs, 154, 18 
\bibitem[Huchra \& Burg(1992)]{HB92} Huchra, J., \& Burg, 
R.\ 1992, \apj, 393, 90 
\bibitem[Imanishi(2002)]{Im02} Imanishi, M.\ 2002, \apj, 
569, 44 
\bibitem[Imanishi(2003)]{Im03} Imanishi, M.\ 2003, \apj, 
599, 918 
\bibitem[Imanishi \& Alonso-Herrero(2004)]{IA04} Imanishi, 
M., \& Alonso-Herrero, A.\ 2004, \apj, 614, 122 
\bibitem[Imanishi et al.(2006)]{ID06} Imanishi, M., Dudley, C. C., \& Maloney, P. R. 2006, ApJ, 637, 114
\bibitem[Imanishi \& Wada(2004)]{IW04} Imanishi, M., \& 
Wada, K.\ 2004, \apj, 617, 214 
\bibitem[Imanishi et al.(2007)]{Im07} Imanishi, M., Dudley, 
C. C., Maiolino, R., Maloney, P. R., Nakagawa, T., \& Risaliti, G.\ 2007, 
\apjs, 171, 72 
\bibitem[Isobe et al.(1986)]{Is86} Isobe, T., Feigelson, 
E. D., \& Nelson, P. I.\ 1986, \apj, 306, 490 
\bibitem[Kaspi et al.(2000)]{Ka00} Kaspi, S., Smith, P. S., 
Netzer, H., Maoz, D., Jannuzi, B. T., \& Giveon, U.\ 2000, \apj, 533, 631 
\bibitem[Kawakatu \& Umemura(2002)]{KU02} Kawakatu, N., \& 
Umemura, M.\ 2002, \mnras, 329, 572
\bibitem[Kawakatu et al.(2007)]{Ka07} Kawakatu, N., 
Imanishi, M., \& Nagao, T.\ 2007, \apj, 661, 660 
\bibitem[Knapen et al.(1995)]{Kn95} Knapen, J. H., Beckman, 
J. E., Heller, C. H., Shlosman, I., \& de Jong, R. S.\ 1995, \apj, 454, 623 
\bibitem[Knapen et al.(2002)]{Kn02} Knapen, J. H., 
P{\'e}rez-Ram{\'{\i}}rez, D., \& Laine, S.\ 2002, \mnras, 337, 808 
\bibitem[Knapen(2005)]{Kn05} Knapen, J. H.\ 2005, \aap, 429, 
141 
\bibitem[Leitherer et al.(1996)]{Le96} Leitherer, C., Vacca, 
W. D., Conti, P. S., Filippenko, A. V., Robert, C., \& Sargent, W. L. W.\ 
1996, \apj, 465, 717 
\bibitem[Lutz et al.(1996)]{Lu96} Lutz, D., et al.\ 1996, 
\aap, 315, L269
\bibitem[Maciejewski et al.(2002)]{Ma02} Maciejewski, W., 
Teuben, P. J., Sparke, L. S., \& Stone, J. M.\ 2002, \mnras, 329, 502 
\bibitem[Makino(1997)]{Ma97} Makino, J.\ 1997, \apj, 478, 58 
\bibitem[Maoz et al.(1996)]{Ma96} Maoz, D., Barth, A. J., 
Sternberg, A., Filippenko, A. V., Ho, L. C., Macchetto, F. D., Rix, H.-W., 
\& Schneider, D. P.\ 1996, \aj, 111, 2248 
\bibitem[Marconi et al.(1994)]{Ma94} Marconi, A., Moorwood, 
A. F. M., Origlia, L., \& Oliva, E.\ 1994, The Messenger, 78, 20 
\bibitem[Matsushita et al.(2005)]{Ma05} Matsushita, S., 
Kawabe, R., Kohno, K., Matsumoto, H., Tsuru, T. G., \& Vila-Vilar{\'o}, B.\ 
2005, \apj, 618, 712
\bibitem[Mauder et al.(1994)]{Mau94} Mauder, W., Weigelt, G., 
Appenzeller, I., \& Wagner, S. J.\ 1994, \aap, 285, 44 
\bibitem[Mihos \& Hernquist(1996)]{MH96} Mihos, J. C., \& 
Hernquist, L.\ 1996, \apj, 464, 641 
\bibitem[Montenegro et al.(1999)]{Mo99} Montenegro, L. E., 
Yuan, C., \& Elmegreen, B. G.\ 1999, \apj, 520, 592 
\bibitem[Mouri et al.(1990)]{Mo90} Mouri, H., Kawara, K., 
Taniguchi, Y., \& Nishida, M.\ 1990, \apjl, 356, L39 
\bibitem[Noguchi(1988)]{No88} Noguchi, M.\ 1988, \aap, 203, 
259 
\bibitem[Norman \& Ikeuchi(1989)]{NI89} Norman, C. A., \& 
Ikeuchi, S.\ 1989, \apj, 345, 372 
\bibitem[Ohsuga \& Umemura(1999)]{OU99} Ohsuga, K., \& 
Umemura, M.\ 1999, \apjl, 521, L13 
\bibitem[Ohsuga et al.(1999)]{Oh99} Ohsuga, K., Umemura, M., 
Fukue, J., \& Mineshige, S.\ 1999, \pasj, 51, 345
\bibitem[Ohsuga \& Umemura(2001)]{OU01} Ohsuga, K., \& 
Umemura, M.\ 2001, \apj, 559, 157 
\bibitem[O'Neill et al.(2005)]{On05} O'Neill, P. M., Nandra, 
K., Papadakis, I. E., \& Turner, T. J.\ 2005, \mnras, 358, 1405 
\bibitem[Onodera et al.(2004)]{On04} Onodera, S., Koda, J., 
Sofue, Y., \& Kohno, K.\ 2004, \pasj, 56, 439 
\bibitem[Panessa \& Bassani(2002)]{PB02} Panessa, F., \& 
Bassani, L.\ 2002, \aap, 394, 435 
\bibitem[Panessa et al.(2006)]{Pa06} Panessa, F., Bassani, 
L., Cappi, M., Dadina, M., Barcons, X., Carrera, F. J., Ho, L. C., \& 
Iwasawa, K.\ 2006, \aap, 455, 173
\bibitem[Peeters et al.(2004)]{Pe04} Peeters, E., Spoon, 
H. W. W., \& Tielens, A. G. G. M.\ 2004, \apj, 613, 986
\bibitem[Peterson \& Wandel(1999)]{PW99} Peterson, B. M., \& 
Wandel, A.\ 1999, \apjl, 521, L95 
\bibitem[Pogge(1989)]{Po89} Pogge, R. W.\ 1989, \apj, 345, 
730 
\bibitem[Rayner et al.(2003)]{Ra03} Rayner, J. T., Toomey, D. W., Onaka, P. M., Denault, A. J., Stahlberger, W. E., Vacca, W. D., Cushing, M. C., \& Wang, S. 2003, PASP, 115, 362
\bibitem[Rieke \& Lebofsky(1985)]{RL85} Rieke, G. H., \& 
Lebofsky, M. J.\ 1985, \apj, 288, 618 
\bibitem[Rodr{\'{\i}}guez-Ardila \& Viegas(2003)]{RV03} 
Rodr{\'{\i}}guez-Ardila, A., \& Viegas, S. M.\ 2003, \mnras, 340, L33 
\bibitem[Rush et al.(1993)]{Ru93} Rush, B., Malkan, M. A., 
\& Spinoglio, L.\ 1993, \apjs, 89, 1 
\bibitem[Saitoh \& Wada(2004)]{SW04} Saitoh, T. R., \& Wada, 
K.\ 2004, \apjl, 615, L93
\bibitem[Shapiro \& Field(1976)]{SF76} Shapiro, P. R., \& 
Field, G. B.\ 1976, \apj, 205, 762
\bibitem[Shinozaki et al.(2006)]{Sh06} Shinozaki, K., 
Miyaji, T., Ishisaki, Y., Ueda, Y., \& Ogasaka, Y.\ 2006, \aj, 131, 2843 
\bibitem[Shlosman et al.(1990)]{Sh90} Shlosman, I., 
Begelman, M. C., \& Frank, J.\ 1990, \nat, 345, 679 
\bibitem[Siebenmorgen et al.(2004)]{Si04} Siebenmorgen, R., 
Kr{\"u}gel, E., \& Spoon, H. W. W.\ 2004, \aap, 414, 123 
\bibitem[Soifer et al.(2002)]{So02} Soifer, B. T., 
Neugebauer, G., Matthews, K., Egami, E., \& Weinberger, A. J.\ 2002, \aj, 
124, 2980
\bibitem[Storchi-Bergmann et al.(1996)]{St96} 
Storchi-Bergmann, T., Wilson, A. S., \& Baldwin, J. A.\ 1996, \apj, 460, 
252 
\bibitem[Taniguchi(1999)]{Ta99} Taniguchi, Y.\ 1999, \apj, 
524, 65 
\bibitem[Thompson et al.(2005)]{Th05} Thompson, T. A., 
Quataert, E., \& Murray, N.\ 2005, \apj, 630, 167 
\bibitem[Tokunaga(2000)]{To00} Tokunaga, A. T. 2000, in Allen's Astrophysical Quantities, ed. A. N. Cox (4th ed; Berlin: Springer), 143
\bibitem[Tokunaga et al.(1991)]{To91} Tokunaga A. T., Sellgren K., Smith R. G., Nagata T., Sakata A., Nakada Y., 1991, ApJ, 380, 452 
\bibitem[Tomisaka \& Ikeuchi(1986)]{TI86} Tomisaka, K., \& 
Ikeuchi, S.\ 1986, \pasj, 38, 697
\bibitem[Toomre(1964)]{To64} Toomre, A.\ 1964, \apj, 139, 
1217 
\bibitem[Tremaine et al.(2002)]{Tr02} Tremaine, S., et al.\ 
2002, \apj, 574, 740 
\bibitem[Turner et al.(1997)]{Tu97} Turner, T. J., George, 
I. M., Nandra, K., \& Mushotzky, R. F.\ 1997, \apjs, 113, 23 
\bibitem[Umemura et al.(1997)]{Um97} Umemura, M., Fukue, J., 
\& Mineshige, S.\ 1997, \apjl, 479, L97 
\bibitem[Umemura et al.(1998)]{Um98} Umemura, M., Fukue, J., 
\& Mineshige, S.\ 1998, \mnras, 299, 1123 
\bibitem[Voit(1992)]{Vo92} Voit, G. M.\ 1992, \mnras, 258, 
841 
\bibitem[Wada et al.(2002)]{Wa02} Wada, K., Meurer, G., \& 
Norman, C. A.\ 2002, \apj, 577, 197 
\bibitem[Wada \& Norman(1999)]{WN99} Wada, K., \& Norman, 
C. A.\ 1999, \apjl, 516, L13 
\bibitem[Wada \& Norman(2001)]{WN01} Wada, K., \& Norman, 
C. A.\ 2001, \apj, 547, 172 
\bibitem[Wada \& Norman(2002)]{WN02} Wada, K., \& Norman, 
C. A.\ 2002, \apjl, 566, L21 
\bibitem[Watabe \& Umemura(2005)]{WU05} Watabe, Y., \& Umemura, M.\ 2005, \apj, 618, 649
\bibitem[Werner et al.(2004)]{We04} Werner, M. W., et al.\ 
2004, \apjs, 154, 1 
\bibitem[Wilson et al.(1991)]{Wi91} Wilson, A. S., Helfer, 
T. T., Haniff, C. A., \& Ward, M. J.\ 1991, \apj, 381, 79 
\end{thebibliography}
\end{document}